\documentclass[aps,twocolumn,superscriptaddress,raggedbottom,showpacs,floatfix]{revtex4-1}
\usepackage[pdftex]{graphicx}
\usepackage{mmap}
\usepackage{amssymb}
\usepackage{amsmath}
\usepackage{color}
\usepackage{hyperref}
\usepackage{textcomp}
\hypersetup{colorlinks=true,linkcolor=blue,urlcolor=blue,citecolor=blue}

\newcommand{\im}{\mathop{\rm Im}}

\begin{document}
\title{\textbf{Hyperbolic hybrid waves and optical topological transitions
	 \\[0.2em]
	 in few-layer anisotropic metasurfaces}}
\author{Oleg~V. Kotov}%
\email{oleg.v.kotov@yandex.ru}%
\affiliation{N.~L. Dukhov Research Institute of Automatics (VNIIA), 127055 Moscow, Russia}%
\affiliation{Institute of Microelectronics Technology and High Purity Materials RAS, 142432 Chernogolovka, Russia}%
\author{Yurii~E. Lozovik}%
\email{lozovik@isan.troitsk.ru}%
\affiliation{Institute for Spectroscopy RAS, 142190 Troitsk, Moscow, Russia}%
\affiliation{N.~L. Dukhov Research Institute of Automatics (VNIIA), 127055 Moscow, Russia}%
\affiliation{Institute of Microelectronics Technology and High Purity Materials RAS, 142432 Chernogolovka, Russia}%
\affiliation{National Research University Higher School of Economics, 101000 Moscow, Russia}%

\begin{abstract}
	A comprehensive analysis of hybrid TM-TE polarized surface electromagnetic waves supported by different few-layer anisotropic metasurfaces is presented. A generalized 4$\times$4 T-matrix formalism for arbitrary anisotropic 2D layers is developed, from which the general relations for the surface waves dispersions and scattering coefficients are deduced. Using this formalism and the effective conductivity approach, the dispersions and iso-frequency contours (IFCs) topology of the surface waves in various hybrid uniaxial metasurfaces are studied. The existence of \textit{hyperbolic plasmon-exciton polaritons} in plasmon-exciton hybrids and \textit{hyperbolic acoustic waves} with strong confinement in both out-of-plane and in-plane directions in uniaxial plasmonic bilayers are predicted. In plasmonic uniaxial metasurfaces on metal films, the \textit{elliptic and hyperbolic backward surface waves} with negative group velocity are predicted and \textit{additional topological transitions} in both elliptic and hyperbolic IFCs of the hybrid surface waves are revealed. Ultrathin twisted uniaxial plasmonic bilayers are proposed as systems with the IFCs topological transitions highly sensitive to the layers twist. The developed formalism may become a useful tool in the calculation of multifunctional few-layer metasurfaces or van der Waals heterostructures based on 2D materials with in-plane anisotropy, where the TM-TE polarization mixing must be considered. The predicted effects may open new horizons in the development and applications of planar optical technologies.

\end{abstract}

\maketitle

\section{Introduction} \label{Sec1}
Metasurfaces, the two-dimensional (2D) analog of metamaterials, have recently gained significant attention as a great candidate for the efficient control over surface electromagnetic (EM) waves \cite{MetaYu2011,MetaAlu2011}. Providing efficient beam shaping, phase and polarization manipulation of light they can serve as optical control devices such as polarization transformers, antennas, perfect absorbers, switchers, sensors, frequency selectors etc. \cite{RevMeta_Holloway,RevMeta_Capasso,RevMeta_ITMO,RevMeta_Chen,RevMeta_Krasnok}. In contrast to bulk metamaterials, while retaining similar functionalities, they allow to remove the volumetric losses, to simplify the fabrication process, and to provide a full on-chip incorporation into planar optical devices. Metasurfaces are called \textit{hyperbolic} when, due to extreme in-plane anisotropy, they behave within the sheet as a dielectric along one direction and as a metal along the orthogonal one \cite{Alu_PRL,ITMO_PRB}. Hyperbolic metasurfaces, following their 3D analog \cite{Smith_2003,Jacob_2012,Drachev_2013,RevHMM_ITMO,RevHMM_Jacob}, have attracted great interest owing to their unique EM properties, such as negative refraction, large density of states, surface plasmon polaritons (SPPs) self-collimation, unique SPPs optical spin control, and photoluminescence polarization anisotropy \cite{Alu_Rev,Hanson_PRL,Liu_thermal,Alu2017,HMS_Lum, ITMO_spin,RevHMS_2019}. Depending on the constituent materials and geometrical parameters the in-plane hyperbolic response can be realized in ultraviolet (UV), optical, infrared (IR), THz, and microwave ranges. In particular, silver grating \cite{HMS_Liu2013,Visible_HMS} or gold elliptical nanodisks \cite{ITMO_PRB,ITMO_expOpt} have been proposed for the optical SPPs, nanostructured h-BN for the mid-IR phonon-polaritons \cite{hBN_Nikitin,hBN_Basov}, thin h-BN film, depending on the thickness, for the mid-IR, near-IR, optical, or even UV SPPs \cite{Hanson_BP, Alu_BP, Katsnelson_BP}, graphene grating for the THz SPPs \cite{Alu_PRL,Alu_OME,Alu_Nonloc,HMS_Jiang}, and anisotropic metallic patterns with centimeter-sized unit cells for the microwave SPPs \cite{ITMO_expGHz, GHz_Yang, GHz_YangMagnet, mushroom}. However, high ohmic losses in metallic structures can sufficiently spoil the outstanding properties of hyperbolic surface waves \cite{Alu_Rev}. Perhaps, all-dielectric metasurfaces with hyperbolic-like regimes for the waveguide (WG) modes \cite{ITMO_all-diell} may become one of the solutions of this problem.   

In order to make metasurfaces active and highly tunable, semiconducting constituent materials can be used. One way is to design thin metal-semiconductor nanostructures based on distributed semiconducting quantum wells \cite{HMS_Lum}. But the most natural way is just to combine already prepared plasmonic metasurfaces with organic dye molecules \cite{Vasa2010,Vasa2013,Ramezani}, ordinary quantum wells \cite{Vasa2008}, or 2D semiconductors \cite{Agarwal,Ding}. Among all 2D semiconductors, monolayers of transition metal dichalcogenides (TMDCs) have aroused large interest for the past few years owing to their direct band gap in the visible and near-IR ranges, valley-selective response, large exciton binding energy and oscillator strengths, high emission quantum yield and strong photoluminescence \cite{RevTMD_2012,RevTMD_2016,RevTMD_Stern,RevTMD_Glazov,RevTMD_Krasnok}. TMDCs also have a number of technological advantages, including, high epitaxial quality, chemical and thermal stability, compatibility with other materials, and relatively abundance \cite{RevTMD_Duan,RevTMD_Brent,RevTMD_Yazyev,RevTMD_Usp}. All this makes TMDCs one of the best candidates for the semiconducting component of hybrid plasmon-exciton metasurfaces with a strong-coupling regime at room temperature \cite{RevTMD_Krasnok,Baranov_Rev,Wen_TMD,Baranov2017,Baumberg,Urbaszek,Baranov2018, Baranov2019}. Such hybrids supporting plasmon-exciton polaritons (plexcitons \cite{Halas2008}) may become a modern platform for ultrafast active control of light \cite{Vasa2010,Vasa2013,Ebbesen2011} and room temperature polariton lasing \cite{Plex_laser}.  

Over the past few years, the 2D materials research has grown into the broader field, which includes the study of van der Waals heterostructures \cite{Geim2013} and transdimensional materials \cite{Boltasseva}. Similar to van der Waals materials, which for some applications are more effective in a few-layer configuration than in a monolayer one \cite{Baranov2018,Baranov2019,Rev_BP}, the applicability of metasurfaces can be dramatically improved by going to the few layers. The multiplication of metasurface layers, while retaining a relatively low level of losses and fabrication simplicity, can provide higher efficiency and more degrees of freedom for manipulating the phase, amplitude, polarization, propagation, and dispersion of light \cite{Few_Rev2015,bi_Zhou}. Moreover, the layers interaction results in additional effects, including near-field coupling, WG modes, and multiple wave interference \cite{Few_Rev2019}. The WG effects enable us to control polarization and phase of the transmitted light simultaneously \cite{WG_Li}, while multiple interference effects can be used to cancel the undesired light and enhance the efficiency of antireflection coatings \cite{Interf2010}, polarization converters \cite{Interf2013,Interf2016}, and metalenses \cite{Interf2017}. Near-field coupling generates a magnetic resonance inside the structure, which being effectively coupled to an in-layer electric resonance allows us to realize perfect absorbers and reflectors \cite{Bozhevolnyi_Rev,Abs_Rev1,Abs_Rev2}. Such a magnetic resonance mode is also named as gap-plasmon mode \cite{Bozhevolnyi_Rev}, which in the case of antisymmetric field profile corresponds to ultraconfined acoustic plasmons \cite{Ac_Koppens,Ac_Lee,Ac_BP}. 

Recently, the research of ultrathin active chiral metamaterials has given rise to the topic of moir\'{e} metasurfaces, which are stacks of two or more periodic patterns with relative differences in lattice constants or in-plane rotation angles \cite{Moire_Rev,Moire2009,Moire2016,Moire2018}. The optical response of such metasurfaces is ultrasensitive to the layers twist or interlayer refractive index \cite{Moire2018}. On the other side, the interest in moir\'{e} structures stirred up after the recent discovery of exotic strongly correlated 
quantum phases in twisted bilayer graphene, such as unconventional insulating \cite{cao2018_Ins} and superconducting phases \cite{cao2018_Sc}, which are highly sensitive to the layers twist angle near the magic values. All this has attracted interest to twisted bilayer plasmonics with moir\'{e} pattern \cite{Tw_Moon,Tw_Stauber2013,Tw_Stauber2016,Tw_Basov,Tw_Polini,Tw_Catarina} and without it \cite{Tw_Renuka,Tw_Ge}, as well as to twisted bilayer excitonics \cite{Twex_tran2019,Twex_seyler2019}.

Theoretical description of few-layer metasurfaces optical properties can be divided into two main problems: homogenization of individual layers and EM scattering of multilayer stack formed by these layers. The first can be done using various homogenization methods \cite{Belov2005,Alu2011} that extract effective parameters from the scattering properties of interacting meta-atoms in the long-wavelength regime, when the effective wavelengths and averaged field variations are much larger than the material granularity. For 2D metasurfaces with periodicity $L$, such homogenization procedures give the description within an effective conductivity approach  \cite{Alu_PRL,Alu_OME,Alu_Nonloc,Tretyakov_book,Luukkonen,ITMO_cond}, which applicable at $L\ll\lambda_{\rm SPP}$. The EM scattering problem of periodic multilayer stacks consisting of 2D isotropic nonmagnetic layers separated by dielectric slabs can be solved analytically using 2$\times$2 transfer-matrix (T-matrix) formalism \cite{jones1941,abeles1950,yeh1977}, which was applied for graphene multilayers in numerous papers \cite{T_Zhan,T_Bludov,T_Iorsh,T_MacDonald,T_Fan,T_Smirnova,T_Deng}. This method allows us to obtain collective evanescent or WG modes dispersions and reflection/transmission coefficients of a multilayer structure. To analyze SPPs in-plane field distribution, it is convenient to consider their dispersions in terms of iso-frequency contours (IFCs), which topology indicates the peculiarities of SPPs propagation, e.g., the switching between omnidirectional and collimation regimes \cite{Alu_PRL,ITMO_PRB}. In metamaterial analyses, IFC plays the role of the Fermi surface in a metal, so IFCs topological transitions from a closed to an open geometry can be called the optical analog of Lifshitz transitions for the Fermi surface, or the optical topological transitions \cite{Jacob_2012}. For the sake of brevity, we further omit the word \textquotedblleft{optical}\textquotedblright and imply a change not in some topological invariants but only in the IFCs topology. When the anisotropy of constituent 2D layers is crucial, or in the presence of an external magnetic field, to account for the TM-TE polarization mixing, the 4$\times$4 T-matrix or S-matrix formalism should be used \cite{teitler1970,berreman1972,yeh1979,S-Inampudi,S-Menzel,T4_Ardakani,S-Achouri}. To account for the magnetoelectric coupling in bianisotropic metasurfaces \cite{Rev_bianiso}, it is more convenient to use the generally accepted in microwave optics impedance matrix formalism and T-circuit representation \cite{Tretyakov_book}.  

In this work, we develop a generalized 4$\times$4 T-matrix formalism allowing to calculate the linear optical response of multilayer metasurfaces consisting of arbitrary anisotropic 2D layers and accounting for the TM-TE polarization mixing, which is critically important for the hyperbolic waves calculations. Using this formalism and the effective conductivity approach we analytically obtain a general dispersion relation for an arbitrary bilayer metasurface. We analyze the dispersions and IFCs topology of the hybrid waves in various few-layer anisotropic metasurfaces in the most general form, not specifying a design of constituent 2D layers. Having considered four examples of hybrid uniaxial metasurfaces, we predict in them the existence of different hybrid hyperbolic waves and additional topological transitions. 

The rest of the paper is organized as follows. In Sec.~\ref{Sec2}, a generalized 4$\times$4 T-matrix formalism for arbitrary anisotropic 2D layers is developed, from which the general relations for the surface waves dispersions and reflection/transmission coefficients are deduced. In Sec.~\ref{Sec3}, four different examples of hybrid uniaxial metasurfaces are considered: hyperbolic plasmon-exciton metasurfaces (\ref{Subs3a}), plasmonic uniaxial metasurfaces on metal or dielectric films (\ref{Subs3b}), bilayer hyperbolic metasurfaces (\ref{Subs3c}), and twisted bilayer hyperbolic metasurfaces (\ref{Subs3d}). In Sec.~\ref{Sec4}, the results are summarized.

\section{Transfer matrix formalism} \label{Sec2}
\subsection{Monolayer metasurface} \label{Subs2a}
\begin{figure}[htp]
	\centering
	\includegraphics[width=1\columnwidth]{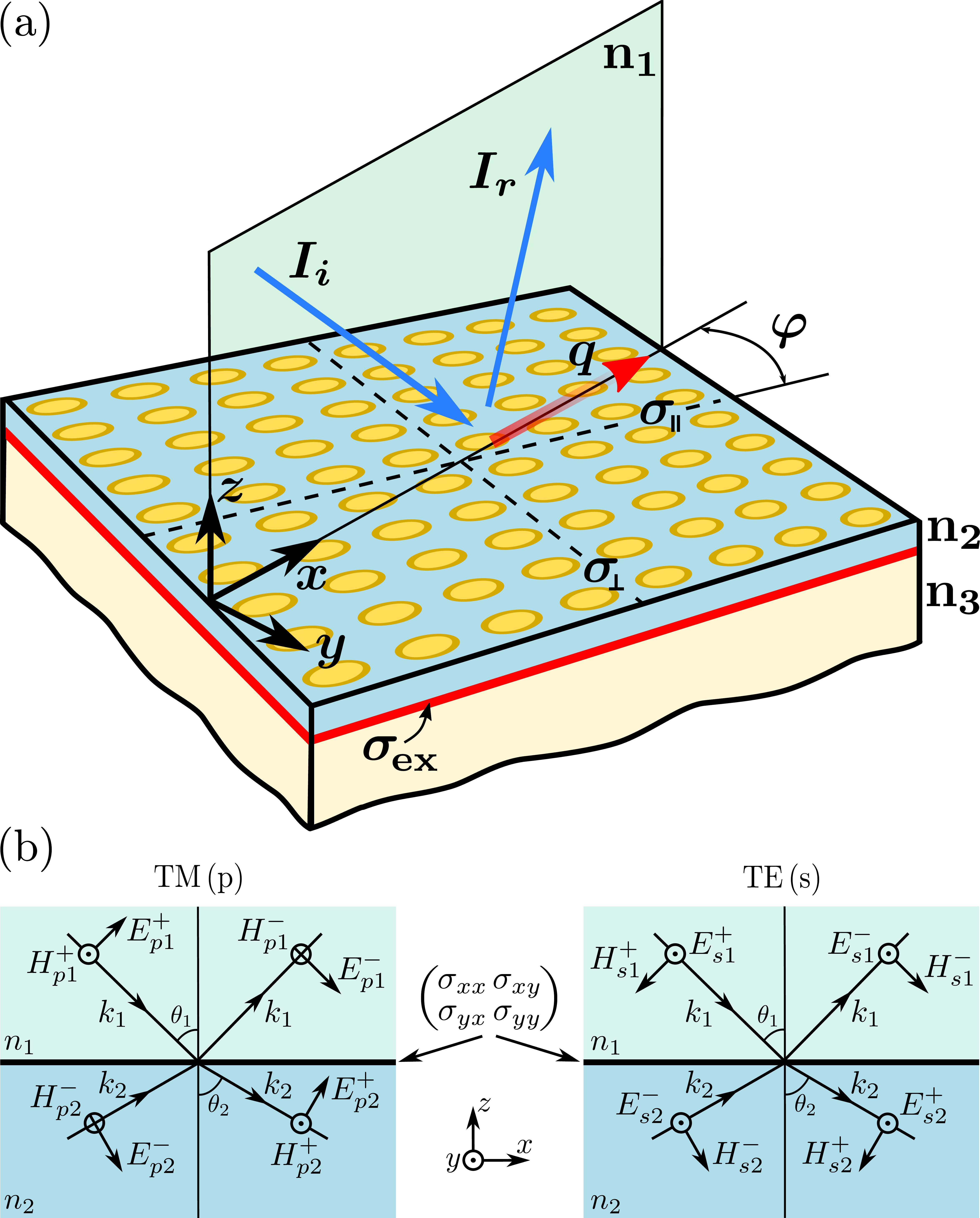} 
	\caption{\label{Fig1} (a) Schematic representation of a light scattering at the plasmon-exciton hybrid metasurface consisting of 2D uniaxial plasmonic array with effective conductivities along ($\sigma_{\parallel}$) and across ($\sigma_{\perp}$) the main axis, thin spacer with refractive index $n_2$, and 2D excitonic layer with isotropic conductivity $\sigma_{\rm ex}$. The refractive indexes of the media above and below the hybrid metasurface are $n_1$ and $n_3$, respectively. The wave vector of the surface waves $q\equiv k_{x}$ is denoted by the red arrow. The plane of incidence of light is at an angle $\varphi$ to the main axis of uniaxial plasmonic array, which gives mixing of the incident light polarizations in the metasurface optical response.  (b) EM field vectors in media above ($n_1$) and below ($n_2$) single anisotropic metasurface for both polarizations, which are mixed by a fully populated conductivity tensor $\widehat\sigma$ of the metasurface.} 
\end{figure}

Let us first consider a single anisotropic metasurface at the interface between two semi-infinite media with refractive indexes $n_1=\sqrt{\varepsilon_1\mu_1}$ and $n_2=\sqrt{\varepsilon_2\mu_2}$ [see Fig.~\ref{Fig1}(b)]. Within the homogenization approach, the EM response of such a metasurface, in general, can be described by a fully-populated conductivity tensor 
\begin{align}
\widehat\sigma =\begin{pmatrix}
\sigma_{xx} & \sigma_{xy} \\[0.5em]  \sigma_{yx} & \sigma_{yy}
\end{pmatrix}.
\end{align}	
Following Refs.~\cite{Nakayama,СhiuQuinn} let us write separately the EM field of the p-polarized (TM) and s-polarized (TE) components of the EM wave, which then will be mixed by the nondiagonal response $\sigma_{xy}$ of a metasurface. For the EM waves with the plane of incidence \textit{xz} [see Fig.~\ref{Fig1}(a)] the wave vectors in the media above ($j=1$) and below ($j=2$) the metasurface are $k_j=n_j\omega/c=\sqrt{k_{xj}^2+k_{zj}^2}$, where $\omega$ is the radiation frequency and $k_{xj}\equiv q$ is the wave vector of the surface waves. The p-waves with the magnetic filed perpendicular to the plane of incidence possess the EM field components  $\textbf{E}_p=\left\lbrace E_x,0,E_z\right\rbrace$, $\textbf{H}_p=\left\lbrace 0,H_y,0\right\rbrace$. For the  angles of incidence $\theta_j$ in the corresponding medium, the projection factors $\cos\theta_j=ck_{zj}\big/\!\!\left(\omega n_j\right)$, which has the same sign for the both forward ($E_{pj}^{+}$) and backward ($E_{pj}^{-}$) waves [see Fig.~\ref{Fig1}(b)], must be taken into account. Thus, using the relation $H_y=\displaystyle\frac{\omega\varepsilon}{ik_z^2c}\frac{\partial{E_x}}{\partial{z}}$ following from the Maxwell equations for the plane monochromatic p-waves, one can write $E_x$ and $H_y$ in the form
\begin{align}
	E_{xj}(\mathbf{r},t)&=\frac{k_{zj}}{n_j\omega/c}\left[E_{pj}^{+}e^{ik_{zj}z}+E_{pj}^{-}e^{-ik_{zj}z}\right]e^{iqx-i\omega t},\nonumber
	\\[0.5em]
	H_{yj}(\mathbf{r},t)&=\frac{\varepsilon_j}{n_j}\left[E_{pj}^{+}e^{ik_{zj}z}-E_{pj}^{-}e^{-ik_{zj}z}\right]e^{iqx-i\omega t}. \label{p}
\end{align}
For the s-waves, with the electric field perpendicular to the plane of incidence, the EM field components are $\textbf{E}_s=\left\lbrace 0,E_y,0\right\rbrace$, $\textbf{H}_s=\left\lbrace H_x,0,H_z\right\rbrace$. The Maxwell equations for the monochromatic s-waves yield the relation $H_x=\displaystyle\frac{ic}{\omega\mu}\frac{\partial{E_y}}{\partial{z}}$ allowing to write $E_y$ and $H_x$ in the form  
\begin{align}
E_{yj}(\mathbf{r},t)&=\left[E_{sj}^{+}e^{ik_{zj}z}+E_{sj}^{-}e^{-ik_{zj}z}\right]e^{iqx-i\omega t}, \nonumber
\\[0.5em]
H_{xj}(\mathbf{r},t)&=\frac{-k_{zj}}{\mu_j\omega/c}\left[E_{sj}^{+}e^{ik_{zj}z}-E_{sj}^{-}e^{-ik_{zj}z}\right]e^{iqx-i\omega t}.
 \label{s}
\end{align}
The boundary conditions on the metasurface (at $z=0$) for both p- and s-waves can be formulated as
\begin{equation}
\begin{array}{rll}
\displaystyle E_{x1}=E_{x2}&\equiv E_x(0),
\\[1em]
\displaystyle H_{y2} - H_{y1}&= -4\pi/c\left[ \sigma_{xx}E_x(0)+\sigma_{xy}E_y(0)\right],  
\\[1em]
\displaystyle E_{y1}=E_{y2}&\equiv E_y(0),
\\[1em]
\displaystyle H_{x2} - H_{x1}&=  4\pi/c\left[ \sigma_{yy}E_y(0)+\sigma_{yx}E_x(0)\right].
 \label{bound}
\end{array}
\end{equation}
Substituting the fields (\ref{p}) and (\ref{s}) in the boundary conditions (\ref{bound}) at $z=0$, one obtain the 4$\times$4 T-matrix, which gives the relation between all the electric field components in the media above and below the metasurface:
\begin{align}
\begin{pmatrix}
E_{p1}^{+} \\[0.5em] E_{p1}^{-} \\[0.5em]  E_{s1}^{+} \\[0.5em] E_{s1}^{-}
\end{pmatrix}=\widehat{T}_{1\rightarrow2}
\begin{pmatrix}
E_{p2}^{+} \\[0.5em] E_{p2}^{-} \\[0.5em]  E_{s2}^{+} \\[0.5em] E_{s2}^{-}
\end{pmatrix} \label{ET}
\end{align}	   
with 
\begin{align}
\widehat{T}_{1\rightarrow2}=\frac{1}{2}\begin{bmatrix}
\displaystyle\frac{k_2n_1}{\varepsilon_1n_2}\begin{pmatrix}
P_{12}^{++} & P_{12}^{-+} \\[0.75em]  P_{12}^{--} & P_{12}^{+-}
\end{pmatrix} & \displaystyle\frac{n_1}{\varepsilon_1}\sigma_{xy}\begin{pmatrix} 
1 & 1 \\[0.75em] -1 & -1 \end{pmatrix} \\[2em]  \displaystyle\frac{k_2\mu_1}{k_1n_2}\sigma_{yx}\begin{pmatrix} 
1 & 1 \\[0.75em] -1 & -1 \end{pmatrix} & \displaystyle\frac{\mu_1}{k_1}\begin{pmatrix}
S_{12}^{++} & S_{12}^{-+} \\[0.75em]  S_{12}^{--} & S_{12}^{+-}
\end{pmatrix} 
\end{bmatrix},\label{T12}
\end{align}
where the p-waves components are
\begin{align}
P_{12}^{++}&=\frac{\varepsilon_1}{k_1}+\frac{\varepsilon_2}{k_2}+\sigma_{xx},\quad  P_{12}^{-+}=\frac{\varepsilon_1}{k_1}-\frac{\varepsilon_2}{k_2}+\sigma_{xx}, \nonumber
\\[0.5em]
P_{12}^{--}&=\frac{\varepsilon_1}{k_1}-\frac{\varepsilon_2}{k_2}-\sigma_{xx},\quad  P_{12}^{+-}=\frac{\varepsilon_1}{k_1}+\frac{\varepsilon_2}{k_2}-\sigma_{xx}, \nonumber
\end{align}
and the s-waves components are given by
\begin{align}
S_{12}^{++}&=\frac{k_1}{\mu_1}+\frac{k_2}{\mu_2}+\sigma_{yy},\quad  S_{12}^{-+}=\frac{k_1}{\mu_1}-\frac{k_2}{\mu_2}+\sigma_{yy}, \nonumber
\\[0.5em]
S_{12}^{--}&=\frac{k_1}{\mu_1}-\frac{k_2}{\mu_2}-\sigma_{yy},\quad  S_{12}^{+-}=\frac{k_1}{\mu_1}+\frac{k_2}{\mu_2}-\sigma_{yy}. \nonumber
\end{align}
Here and after $k_j$ denotes $k_{zj}$ normalized to $\omega/c$ and all the conductivity tensor components $\sigma_{ij}$ are normalized to $c/4\pi$. Notice that 2$\times$2 T-matrices, diagonally arranged in the matrix (\ref{T12}) and consisting of the components $P_{12}^{\pm}$ and $S_{12}^{\pm}$, are well-known T-matrices of an isotropic 2D layer for p- and s-waves, respectively \cite{T_Zhan}. In general, for any 4$\times$4 T-matrix that links all the electric field components in a first layer with those in N-layer,  
\begin{align}
	\begin{pmatrix}
		E_{p1}^{+} \\[0.5em] E_{p1}^{-} \\[0.5em]  E_{s1}^{+} \\[0.5em] E_{s1}^{-}
	\end{pmatrix}=\begin{pmatrix}
	T_{11} & T_{12} & T_{13} & T_{14}\\[0.5em]
	T_{21} & T_{22} & T_{23} & T_{24}\\[0.5em]
	T_{31} & T_{32} & T_{33} & T_{34}\\[0.5em]
	T_{41} & T_{42} & T_{43} & T_{44}
\end{pmatrix}
\begin{pmatrix}
	E_{pN}^{+} \\[0.5em] E_{pN}^{-} \\[0.5em]  E_{sN}^{+} \\[0.5em] E_{sN}^{-}
\end{pmatrix},
\end{align}	   
the reflection and transmission coefficients are defined and
expressed in terms of the T-matrix elements as follows (see Ref.~\cite{yeh1979}):
\begin{align}
r_{pp}&=\left.\frac{E_{p1}^{-}}{E_{p1}^{+}}\right|_{E_{s1}^{+}=0}=\frac{T_{21}T_{33}-T_{23}T_{31}}{T_{11}T_{33}-T_{13}T_{31}},\nonumber
\\[0.5em]  
r_{ps}&=\left.\frac{E_{s1}^{-}}{E_{p1}^{+}}\right|_{E_{s1}^{+}=0}=\frac{T_{41}T_{33}-T_{43}T_{31}}{T_{11}T_{33}-T_{13}T_{31}},\nonumber
\\[0.5em] 
r_{sp}&=\left.\frac{E_{p1}^{-}}{E_{s1}^{+}}\right|_{E_{p1}^{+}=0}=\frac{T_{11}T_{23}-T_{13}T_{21}}{T_{11}T_{33}-T_{13}T_{31}},\nonumber
\\[0.5em] 
r_{ss}&=\left.\frac{E_{s1}^{-}}{E_{s1}^{+}}\right|_{E_{p1}^{+}=0}=\frac{T_{11}T_{43}-T_{13}T_{41}}{T_{11}T_{33}-T_{13}T_{31}},\nonumber
\\[0.5em]
t_{pp}&=\left.\frac{E_{pN}^{+}}{E_{p1}^{+}}\right|_{E_{s1}^{+}=0}=\frac{T_{33}}{T_{11}T_{33}-T_{13}T_{31}},\nonumber
\\[0.5em] 
t_{ps}&=\left.\frac{E_{sN}^{+}}{E_{p1}^{+}}\right|_{E_{s1}^{+}=0}=\frac{-T_{31}}{T_{11}T_{33}-T_{13}T_{31}},\nonumber
\\[0.5em] 
t_{sp}&=\left.\frac{E_{pN}^{+}}{E_{s1}^{+}}\right|_{E_{p1}^{+}=0}=\frac{-T_{13}}{T_{11}T_{33}-T_{13}T_{31}},\nonumber
\\[0.5em] 
t_{ss}&=\left.\frac{E_{sN}^{+}}{E_{s1}^{+}}\right|_{E_{p1}^{+}=0}=\frac{T_{11}}{T_{11}T_{33}-T_{13}T_{31}},\label{rrtt}
\end{align}
where the condition $E_{pN}^{-}=E_{sN}^{-}=0$ of no backward waves in the last medium was used.  The energy reflection (transmission) coefficients are the ratio of the Poynting vector of the reflected (transmitted) and the incident waves, and expressed through the amplitude coefficients as
\begin{align}
	R_p&=\left|r_{pp}\right|^2+\left|r_{ps}\right|^2, \quad  T_p=\frac{\mu_1k_2}{\mu_2k_1}\left(\left|t_{pp}\right|^2+\left|t_{ps}\right|^2\right), \nonumber
	\\[0.5em] 
	R_s&=\left|r_{ss}\right|^2+\left|r_{sp}\right|^2, \quad T_s=\frac{\mu_1k_2}{\mu_2k_1}\left(\left|t_{ss}\right|^2+\left|t_{sp}\right|^2\right). \label{RRTT}
\end{align}
The dispersion of the collective surface waves in such N-layer system can be found as zeros of the denominator of the reflection and transmission coefficients:
\begin{align}
	T_{11}T_{33}-T_{13}T_{31}=0.\label{SPP}
\end{align}
For monolayer metasurface from Eqs.~(\ref{T12})-(\ref{rrtt}) we get:
\begin{equation}
\begin{array}{rll}
	r^{12}_{pp}&=\displaystyle\frac{P_{12}^{--}S_{12}^{++}+\sigma_{xy}\sigma_{yx}}{\Delta}, \ \ 
	&r^{12}_{sp}=\displaystyle-2\sqrt{\frac{\varepsilon_1}{\mu_1}}\frac{\sigma_{xy}}{\Delta},
	\\[2em]
	r^{12}_{ps}&=\displaystyle-2\sqrt{\frac{\varepsilon_1}{\mu_1}}\frac{\sigma_{yx}}{\Delta}, 
	&r^{12}_{ss}=\displaystyle\frac{P_{12}^{++}S_{12}^{--}+\sigma_{xy}\sigma_{yx}}{\Delta},
	\\[2em] 
	t^{12}_{pp}&=\displaystyle\frac{2\sqrt{\varepsilon_2\mu_2}}{k_2}\sqrt{\frac{\varepsilon_1}{\mu_1}}\frac{S_{12}^{++}}{\Delta}, 
	&t^{12}_{sp}=\displaystyle\frac{-2\sqrt{\varepsilon_2\mu_2}}{k_2}\frac{k_1}{\mu_1}\frac{\sigma_{xy}}{\Delta}, 
	\\[2em] 
	t^{12}_{ps}&=\displaystyle-2\sqrt{\frac{\varepsilon_1}{\mu_1}}\frac{\sigma_{yx}}{\Delta},
	&t^{12}_{ss}=\displaystyle\frac{2k_1}{\mu_1}\frac{P_{12}^{++}}{\Delta},
\end{array}	\\[0.4em] \label{ppss}
\end{equation}
with $\Delta=P_{12}^{++}S_{12}^{++}-\sigma_{xy}\sigma_{yx}$.
Zeros of $\Delta$ give the dispersion relation of the surface waves in a monolayer metasurface: 
\begin{align}
\left(\frac{\kappa_1}{\mu_1}+\frac{\kappa_2}{\mu_2}-i\sigma_{yy}\right)\left(\frac{\varepsilon_1}{\kappa_1}+\frac{\varepsilon_2}{\kappa_2}+i\sigma_{xx}\right)=\sigma_{xy}\sigma_{yx}, \label{SPP1}
\end{align}
where $\kappa_j=\sqrt{(qc/\omega)^2-\varepsilon_j\mu_j}=ik_j$ are normalized to $\omega/c$ inverse penetration depths of the surface waves into the upper and lower medium. Our general results, the reflection and transmission coefficients (\ref{ppss}) and the dispersion relation (\ref{SPP1}), for a monolayer metasurface surrounded by media with arbitrary $\varepsilon_j$ and $\mu_j$ correspond at $\mu_j=1$ to those obtained in Refs.~\cite{Lakhtakia,Nikitin,Alu_OME,Kotov2017}) and Refs.~\cite{Nakayama,СhiuQuinn}, respectively.

\subsection{Bilayer (multilayer) metasurfaces} \label{Subs2b}	
The formalism developed above can be easily generalized for an arbitrary number of layers by multiplying the T-matrices corresponding to each layer. For a multilayer metasurface consisting of $N$ 2D layers with effective conductivity tensors $\widehat\sigma_j\,\,(j=1, 2,.., N)$, which are at the interface between corresponding media with refractive indexes $n_{j}$ and $n_{j+1}$, the total T-matrix is given by 
\begin{align}
	\widehat{T}_{1\rightarrow N+1}=\widehat{T}_{1\rightarrow2}\widehat{T}_{d_1}\widehat{T}_{2\rightarrow3}\cdots\widehat{T}_{d_{N-1}}\widehat{T}_{N\rightarrow N+1}, \label{TN}
\end{align}
where $\widehat{T}_{j\rightarrow j+1}$ is obtained from $\widehat{T}_{1\rightarrow2}$ [Eq.~(\ref{T12})] by replacing media $n_{1,2}$ with $n_{j,j+1}$ and $\widehat\sigma_1$ with $\widehat\sigma_j$, $\widehat{T}_{d_j}$ are the T-matrices for a light propagating through the interlayers (media between two adjacent 2D layers) with corresponding thicknesses $d_j$ ($j=1, 2,.., N-1$):
\begin{align}
	\widehat{T}_{d_j}=\begin{pmatrix}
		e^{-ik_2d_j} & 0 & 0 & 0\\[0.5em]
		0 & e^{ik_2d_j} & 0 & 0\\[0.5em]
		0 & 0 & e^{-ik_2d_j} & 0\\[0.5em]
		0 & 0 & 0 & e^{ik_2d_j}
	\end{pmatrix}. \label{Td}
\end{align}
T-matrix (\ref{TN}) allows to obtain all necessary characteristics (the reflection and transmission coefficients, the dispersion relation) using general Eqs.~(\ref{rrtt})-(\ref{SPP}).

Now and later let us focus on consideration of different bilayer metasurfaces ($n_1|\widehat\sigma_1|n_2|\widehat\sigma_2|n_3$): two 2D layers with effective conductivity tensors $\widehat\sigma_1$ and $\widehat\sigma_2$, which are separated by an interlayer with refractive index $n_2$ and thickness $d_1\equiv d$ and surrounded by semi-infinite media with refractive indexes $n_1$ and $n_3$. The total T-matrix for such a system is given by
\begin{align}
\widehat{T}_{1\rightarrow3}=\widehat{T}_{1\rightarrow2}\widehat{T}_d\widehat{T}_{2\rightarrow3}, \label{T_bi}
\end{align}
where $\widehat{T}_{2\rightarrow 3}$ is obtained from $\widehat{T}_{1\rightarrow2}$ [Eq.~(\ref{T12})] by replacing media $n_{1,2}$ with $n_{2,3}$ and $\widehat\sigma_1$ with $\widehat\sigma_2$, $\widehat{T}_{d}$ is given by Eq.~(\ref{Td}).
The dispersion of the collective surface waves in such a bilayer system can be found from Eq.~(\ref{SPP}), where the elements of the matrix $\widehat{T}_{1\rightarrow3}$ are substituted:
\vspace{-1em} 
\begin{widetext}
\vspace{-1em} 
\begin{align}
	&\biggl[\kappa_2^2\left(P_{12}^{++}P_{23}^{++}+P_{12}^{-+}P_{23}^{--}e^{-2\kappa_2d}\right)+n_2^2\sigma_1^{xy}\sigma_2^{yx}(1-e^{-2\kappa_2d})\biggr]\!\!	\biggl[n_2^2\left(S_{12}^{++}S_{23}^{++}+S_{12}^{-+}S_{23}^{--}e^{-2\kappa_2d}\right)+\kappa_2^2\sigma_1^{yx}\sigma_2^{xy}(1-e^{-2\kappa_2d})\biggr] \nonumber
	\\[0.5em]
	&=\!\!\biggl[\kappa_2^2\sigma_1^{yx}\left(P_{23}^{++}+P_{23}^{--}e^{-2\kappa_2d}\right) +n_2^2\sigma_2^{yx}\left(S_{12}^{++}-S_{12}^{-+}e^{-2\kappa_2d}\right)\biggr]\!\!\biggl[n_2^2\sigma_1^{xy}\left(S_{23}^{++}+S_{23}^{--}e^{-2\kappa_2d}\right) +\kappa_2^2\sigma_2^{xy}\left(P_{12}^{++}-P_{12}^{-+}e^{-2\kappa_2d}\right)\biggr],\label{SPP2}
\end{align}
\vspace{-1em} 
\end{widetext}
\vspace{-1em} 
with
\begin{align}
	P_{12}^{\pm\pm}&=\frac{\varepsilon_1}{\kappa_1}\pm\frac{\varepsilon_2}{\kappa_2}\pm i\sigma_1^{xx},\quad  P_{23}^{\pm\pm}=\frac{\varepsilon_2}{\kappa_2}\pm\frac{\varepsilon_3}{\kappa_3}\pm i\sigma_2^{xx}, \nonumber
	\\[0.5em]
	S_{12}^{\pm\pm}&=\frac{\kappa_1}{\mu_1}\pm\frac{\kappa_2}{\mu_2}\mp i\sigma_1^{yy},\quad  S_{23}^{\pm\pm}=\frac{\kappa_2}{\mu_2}\pm\frac{\kappa_3}{\mu_3}\mp i\sigma_2^{yy}, \nonumber
\end{align}
and $\kappa_j=\sqrt{(qc/\omega)^2-\varepsilon_j\mu_j}$. This general bilayer dispersion, as well as the 4$\times$4 T-matrix (\ref{T12}), are the main analytical results of the paper. In the next section we will consider various realizations of anisotropic bilayer metasurfaces, for which the application of the general relation (\ref{SPP2}) will be demonstrated in some special cases.
\vspace{0.5em} 
\section{Hybrid uniaxial metasurfaces} \label{Sec3}
\subsection{Hyperbolic plasmon-exciton metasurfaces} \label{Subs3a}
Let us consider the plasmon-exciton hybrid metasurface [see Fig.~\ref{Fig1}(a)] consisting of 2D uniaxial plasmonic array characterized by conductivity tensor $\widehat\sigma_1\equiv\widehat\sigma$ and 2D excitonic layer (e.g., 2D semiconductor or dye molecules) described by isotropic conductivity:  $\sigma_2^{xx}=\sigma_2^{yy}=\sigma_{\rm ex}$  and $\sigma_2^{xy}=\sigma_2^{yx}=0$. In this case one can simplify the dispersion Eq.~(\ref{SPP2}) to the form
\begin{widetext}
	\vspace{-1em} 
	\begin{align}
		\frac{\left(P_{12}^{++}P_{23}^{++}+P_{12}^{-+}P_{23}^{--}e^{-2\kappa_2d}\right)\left(S_{12}^{++}S_{23}^{++}+S_{12}^{-+}S_{23}^{--}e^{-2\kappa_2d}\right)}{\left(P_{23}^{++}+P_{23}^{--}e^{-2\kappa_2d}\right)\left(S_{23}^{++}+S_{23}^{--}e^{-2\kappa_2d}\right)}=\sigma_{xy}\sigma_{yx}. \label{SPPexpl}
	\end{align} 
\end{widetext}
In the case when there is no interlayer between 2D layers ($d=0$), one can reduce the general bilayer dispersion Eq.~(\ref{SPP2}) to the form
\begin{align}
\left(\frac{\kappa_1}{\mu_1}+\frac{\kappa_3}{\mu_3}-i\sigma_\Sigma^{yy}\right)\left(\frac{\varepsilon_1}{\kappa_1}+\frac{\varepsilon_3}{\kappa_3}+i\sigma_\Sigma^{xx}\right) =\sigma_\Sigma^{xy}\sigma_\Sigma^{yx}, \label{SPPd0}
\end{align}
where $\widehat\sigma_\Sigma=\widehat\sigma_1+\widehat\sigma_2$. Thus, for a bilayer with $d=0$ we get the monolayer-like dispersion Eq.~(\ref{SPP1}) but with the total effective conductivity tensor consisting of both layers contributions, which is consistent with a naive intuition.

Within the homogenization procedure, which depends on the constituent materials and geometry, one can describe a nonmagnetic achiral 2D uniaxial plasmonic layer by effective conductivities along ($\sigma_{\parallel}$) and across ($\sigma_{\perp}$) the main axis. For a light with the plane of incidence is at an angle $\varphi$ to the main axis, the rotated conductivity tensor should be used \cite{ITMO_PRB,Alu_OME}:

\begin{align}
	&\widehat\sigma =\begin{pmatrix}
	\sigma_\parallel & 0 \\[0.5em] 0 & \sigma_\perp
	\end{pmatrix}_{\varphi} = \nonumber
	\\[0.5em]
	&\begin{pmatrix}
	\sigma_{\parallel}\cos^2{\varphi} +\sigma_{\perp}\sin^2{\varphi} & \left(\sigma_{\perp} -\sigma_{\parallel}\right)\sin{2\varphi}/2 \\[0.5em] \left(\sigma_{\perp} -\sigma_{\parallel}\right)\sin{2\varphi}/2 &  \sigma_{\parallel}\sin^2{\varphi} +\sigma_{\perp}\cos^2{\varphi}
	\end{pmatrix}. \label{sig}
\end{align}
The nondiagonal response $\sigma_{xy}=\left(\sigma_{\perp} -\sigma_{\parallel}\right)\sin{2\varphi}/2$, mixing p- and s-waves, arises here not due to an intrinsic chirality or magnetism of plasmonic layer, but follows only from a nonzero tilt of the plane of incidence of light with respect to the main axis, which corresponds to a so-called \textit{extrinsic} chirality \cite{Plum,Kotov2017}. In the dipole and local response approximations the effective conductivities, describing the resonant interaction between the individual scatterers in a plasmonic layer, can be written in a general Lorentzian form:
\begin{align}
	\displaystyle\sigma_{\parallel,\perp}=\sigma_{\parallel,\perp}^\infty+\frac{A_{\parallel,\perp}i\omega}{\omega^2-\Omega_{\parallel,\perp}^2+i\omega\gamma_{\parallel,\perp}}, \label{sig_pl}
\end{align}	
where $\Omega_{\parallel,\perp}$ and $\gamma_{\parallel,\perp}$ are the resonant frequencies and corresponding bandwidths along and across the main axis, $A_{\parallel,\perp}$ are the corresponding oscillator strengths and $\sigma_{\parallel,\perp}^\infty$ are the corresponding background conductivities caused by a nondipole response or finite thickness of a plasmonic layer. In Fig.~\ref{Fig2}(a) we plot the dimensionless conductivities (\ref{sig_pl}) with the realistic parameters corresponding to a thin ($\sim20\,\rm nm$) plasmonic array (like in Ref.~\cite{ITMO_cond}). Two Lorentzians with different resonant frequencies lead to the three different regimes in such a uniaxial plasmonic array \cite{ITMO_PRB}: at low frequencies $\omega<\Omega_{\parallel}$ a capacitive one when both $\im \sigma_{\perp}$ and $\im \sigma_{\parallel}$ are negative, between the resonant frequencies $\Omega_{\parallel}<\omega<\Omega_{\perp}$ a hyperbolic one when they have different signs, and at high frequencies $\omega>\Omega_{\perp}$ an inductive regime when they are both positive. In the capacitive and inductive regimes the structure supports conventional TM and TE SPPs, respectively, with the elliptic topology of the iso-frequency contours. However, in the hyperbolic regime there are mixed TE-TM SPPs with the hyperbolic topology, which allows to achieve extremely large in-plane field confinement of SPPs in some specific directions \cite{Alu_PRL}. The conductivity of the generic excitonic layer also can be written in a Lorentzian form:
\begin{figure}[htp]
	\centering
	\includegraphics[width=1\columnwidth]{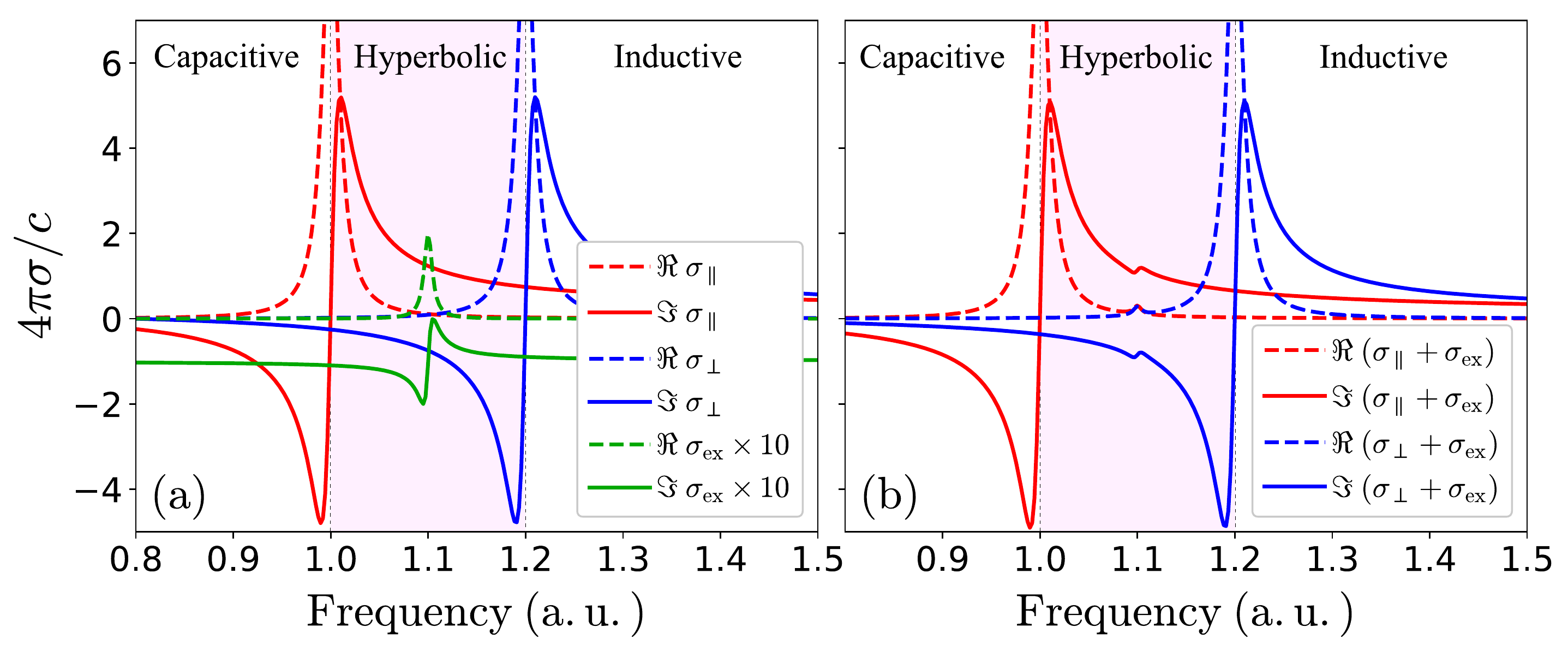} 
	\caption{\label{Fig2} (a) The effective conductivities [Eq.~(\ref{sig_pl})] along ($\sigma_{\parallel}$ red lines) and across ($\sigma_{\perp}$ blue lines) the main axis of the model thin ($\sim20\,\rm nm$) uniaxial plasmonic array, and the ten times magnified conductivity [Eq.~(\ref{sig_ex})] of the $\rm WS_2$-like excitonic layer (green lines). (b) The total parallel and perpendicular effective conductivities of the plasmon-exciton hybrid metasurface with no interlayer between 2D layers. The shaded regions denote the hyperbolic regime. The parameters of the plasmonic array in arbitrary units (a.u.) are $\sigma_{\parallel}^\infty=\sigma_{\perp}^\infty=0.2i$, $A_{\parallel}=A_{\perp}=0.2$, $\gamma_{\parallel}=\gamma_{\perp}=0.02$, $\Omega_{\parallel}=1$, $\Omega_{\perp}=1.2$. The parameters of the $\rm WS_2$-like excitonic layer in a.u. are $\sigma_{\rm ex}^\infty=-0.1i$, $A_{\rm ex}=0.002$, $\Omega_{\rm ex}=1.1$, $\gamma_{\rm ex}=0.01$.} 
\end{figure}
\begin{figure*}[htp]
	\centering
	\includegraphics[width=\textwidth]{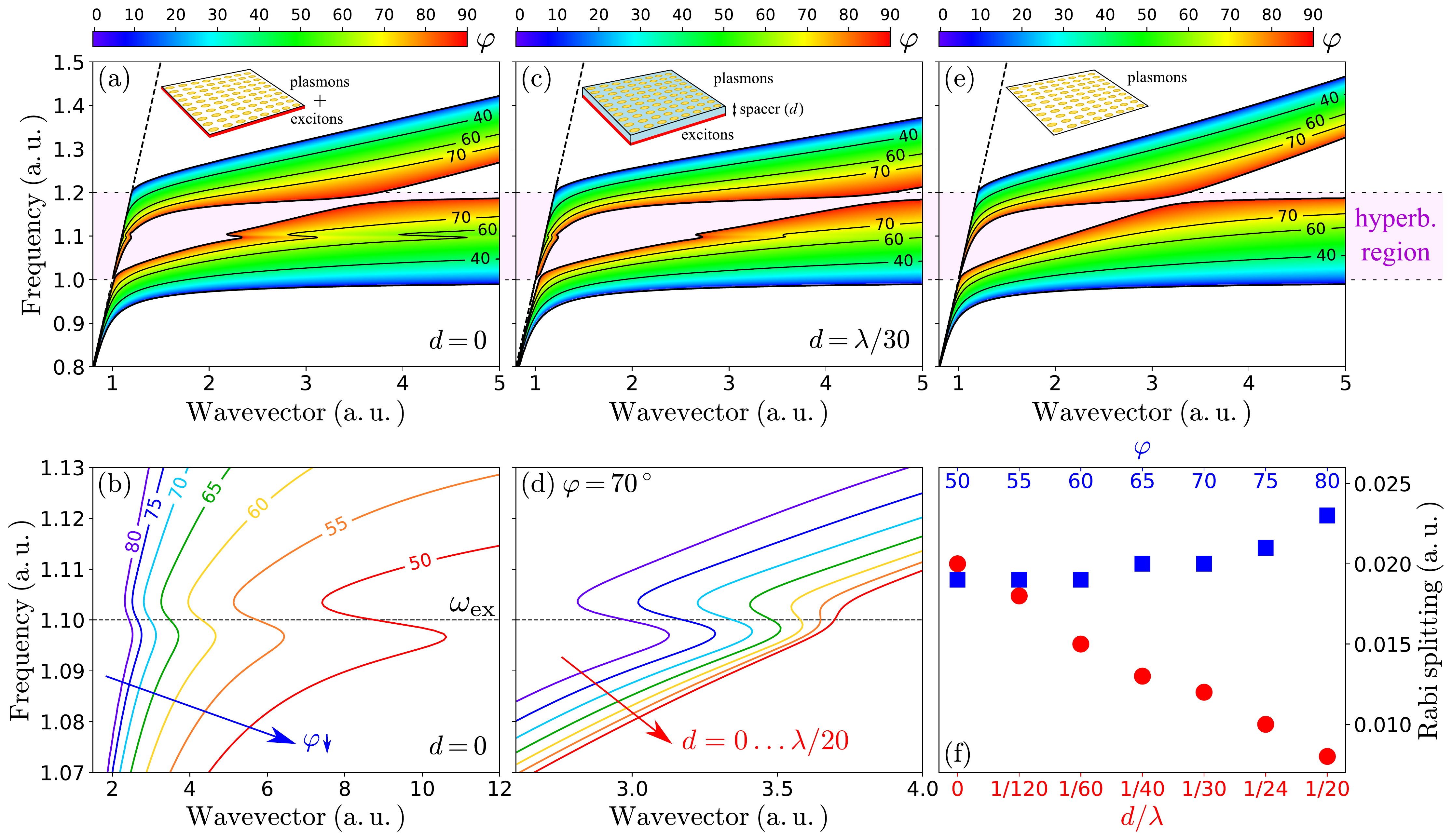} 
	\caption{\label{Fig3} The dispersions $\omega(q)$ of hybrid surface waves at different propagation directions $\varphi$ [see Fig.~\ref{Fig1}(a)] in the plasmon-exciton hybrid metasurface without interlayer between 2D layers [$d=0$] (a) and with it [$\varepsilon_2=2,\,\mu_2=1,\,d=\lambda/30$, $\lambda=2\pi c/\omega$] (c), and in the free-standing plasmonic array [$\sigma_{\rm ex}=0,\,d=0$] (e). The hyperbolic region lying at $\omega\in[\Omega_{\parallel},\Omega_{\perp}]$ is the same for the cases (a),(c), and (e) and marked as a unified shaded bar. The dispersions of the hyperbolic plasmon-exciton polaritons for the set of the angles $\varphi$ at $d=0$ (b) and for the set of the spacer thicknesses $d$ at fixed $\varphi=70^\circ$ (d). (f) The Rabi splitting values for the set of $\varphi$ from (b) [blue squares] and the set of $d$ from (d) [red circles]. All the structures are considered in a free space with $n_1=n_3=1$. The parameters of the plasmonic and excitonic layers are the same as in Fig.~\ref{Fig2}.} 
\end{figure*}
\begin{figure*}[htp]
	\centering
	\includegraphics[width=0.9\textwidth]{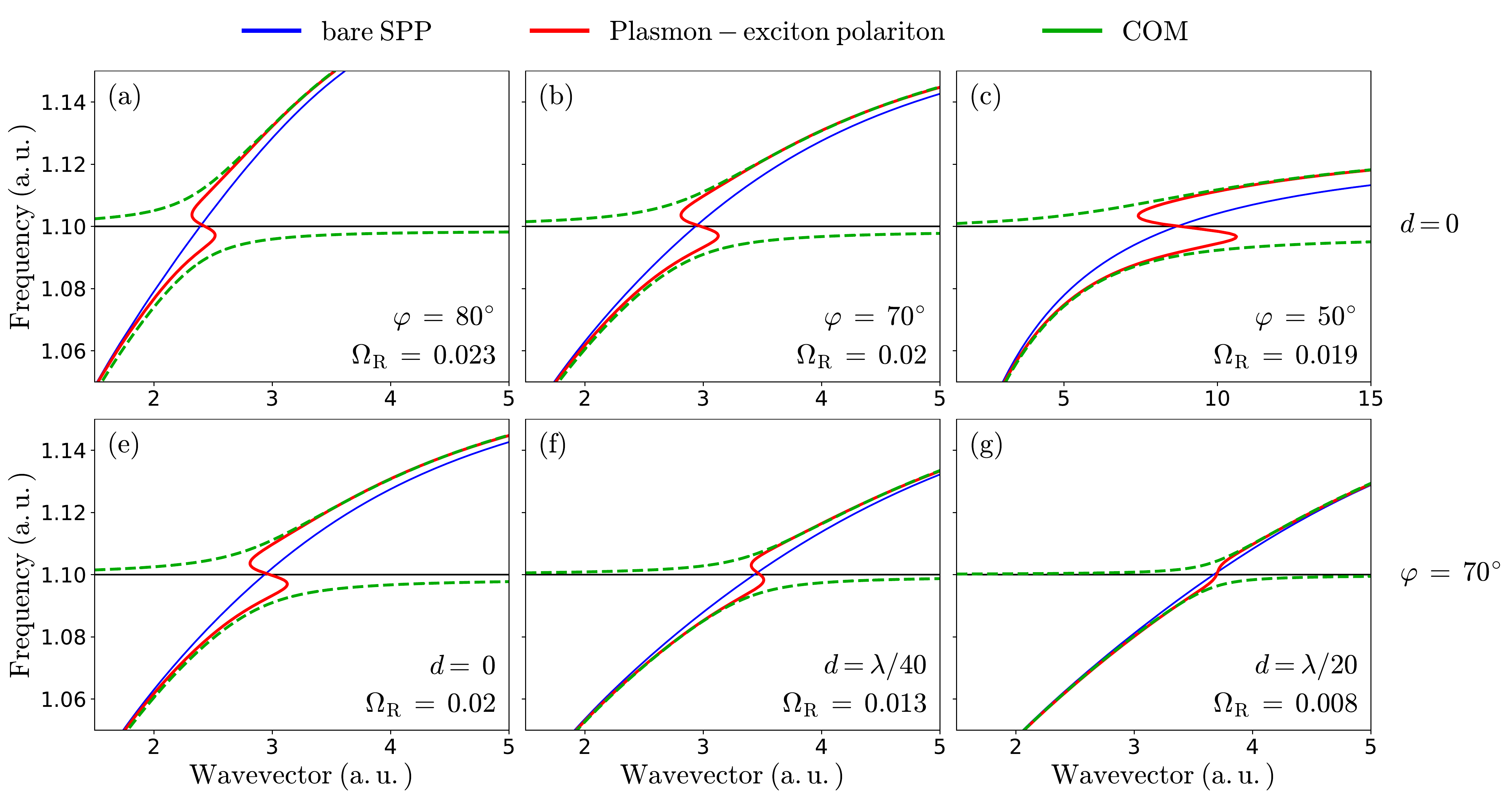} 
	\caption{\label{Fig4} The dispersions $\omega(q)$ of the hyperbolic plasmon-exciton polaritons (red solid lines) in the plasmon-exciton hybrid metasurface and the dispersions of the bare hyperbolic SPPs (hybrid TM-TE waves) [blue solid lines] in the plasmonic array on the thin ($d$) substrate with the spacer parameters ($\varepsilon_2=2,\,\mu_2=1$) at $d=0$ for the angles $\varphi=80^\circ$ (a), $70^\circ$ (b), and $50^\circ$ (c), and at fixed $\varphi=70^\circ$ for the spacer thicknesses $d=0$ (d),  $d=\lambda/40$ (e), and  $d=\lambda/20$ (f), where $\lambda=2\pi c/\omega$. Black solid line denotes the exciton frequency. The retrieved Rabi splitting values are shown in each panel and were used in the COM analyses, which give split polariton branches (green dashed lines) tending to the exact solution when moving away from splitting. Other parameters are the same as for Fig.~\ref{Fig3}.} 
\end{figure*}
\begin{align}
\displaystyle\sigma_{\rm ex}=\sigma_{\rm ex}^\infty+\frac{A_{\rm ex}i\omega}{\omega^2-\Omega_{\rm ex}^2+i\omega\gamma_{\rm ex}}, \label{sig_ex}
\end{align}	
where $\Omega_{\rm ex}$ and  $\gamma_{\rm ex}$ are the excitonic transition frequency and corresponding damping rate, $A_{\rm ex}$ is the oscillator strengths, and $\sigma_{\rm ex}^\infty$ is the background conductivity accounting for the lower electronic bands contributions. In Fig.~\ref{Fig2}(a) we depict the dimensionless conductivity (shown magnified for visibility) with the parameters as for the monolayer TMDC ($\rm WS_2$) \cite{PRB_TMDC}, and $\Omega_{\rm ex}$ lying between $\Omega_{\parallel}$ and $\Omega_{\perp}$, where the plasmonic layer supports the hyperbolic regime. Fig.~\ref{Fig2}(b) shows the total parallel and perpendicular effective conductivities of the plasmon-exciton hybrid metasurface without interlayer between 2D layers [see Eq.~(\ref{SPPd0})]. Since $\im\sigma_{\rm ex}\ll\im\sigma_{\parallel,\perp}$, all regimes (including the hyperbolic one) in the plasmon-exciton hybrid remains the same as in the plasmonic layer.    
 
Using Eq.~(\ref{SPPexpl}) with the conductivities shown in Fig.~\ref{Fig2}, we calculate at $n_1=n_3=1$ the dispersion $\omega(q)$ of the surface waves at different propagation directions (defined by the angles $\varphi$ [see Fig.~\ref{Fig1}(a)]) for the following configurations: free-standing plasmonic array ($\sigma_{\rm ex}=0,\,d=0$), plasmon-exciton hybrid without interlayer between 2D layers ($d=0$), and plasmon-exciton hybrid with thin dielectric spacer ($\varepsilon_2=2,\,\mu_2=1,\,d=\lambda/30$ with $\lambda=2\pi c/\omega$). For the free-standing plasmonic array we obtain the same dispersions of hybrid TM-TE waves as in Ref.~\cite{ITMO_PRB} but with greater anticrossing gap in the hyperbolic region due to the larger bandwidths $\gamma_{\parallel,\perp}$ in the conductivity [see Fig.~\ref{Fig3}(e)]. In the plasmon-exciton hybrid metasurface without interlayer we get a pronounced \textit{hyperbolic plasmon-exciton polaritons} (HPEPs) near the exciton frequency, which lying in the hyperbolic region of the plasmonic layer [see Fig.~\ref{Fig3}(a)]. The dispersion of the HPEPs strongly depends on the propagation direction angle $\varphi$ [see Fig.~\ref{Fig3}(b)]. When there is a thin dielectric spacer between plasmonic and excitonic layers [see Fig.~\ref{Fig3}(c)], the HPEPs are not so pronounced, because HPEPs dispersion loses the excitonic contribution with the spacer thickness increasing [see Fig.~\ref{Fig3}(d)]. Using the obtained dispersions of HPEPs in plasmon-exciton hybrid, the dispersions of the bare SPPs (hybrid TM-TE waves) in the plasmonic array on the substrate with the interlayer parameters ($n_2$, $d$), and standard coupled oscillator model (COM) as in Ref~\cite{Mortensen}, we retrieve Rabi splitting ($\Omega_{\rm R}$) of the plasmon-exciton interaction in the considered hybrid metasurfaces [see Fig.~\ref{Fig4}]. According to the COM (without damping for the simplicity) the two polariton modes frequencies are given by
\begin{align}
\omega_{\rm pol}=\frac{\omega_{\rm SPP}(q)+\Omega_{\rm ex}}{2}\pm\frac{1}{2}\sqrt{\left[\omega_{\rm SPP}(q)-\Omega_{\rm ex}\right]^2+\Omega_{\rm R}^2}\,,\nonumber
\end{align}
where $\omega_{\rm SPP}(q)$ is the bare SPPs (hybrid TM-TE waves) dispersion in the plasmonic array on the substrate with the interlayer parameters and $\Omega_{\rm ex}$ is the exciton frequency from Eq.~(\ref{sig_ex}). We get that the Rabi splitting slowly grows with increasing deflection ($\varphi$) of the propagation direction from the main axis of plasmonic layer [see results at $d=0$ in Fig.~\ref{Fig4}(a)]. We also obtain noticeable decrease of the Rabi splitting with the spacer thickness increasing [see results at fixed  $\varphi=70^\circ$ in Fig.~\ref{Fig4}(b)]. This was to be expected due to the weakening of the plasmon-exciton interaction with increasing of the distance between interacting layers. The Rabi splitting values for the discrete set of the angles from Fig.~\ref{Fig3}(b) and the set of the  thicknesses from Fig.~\ref{Fig3}(d) are gathered in Fig.~\ref{Fig3}(f). It is seen that the Rabi splitting drops with the spacer thickness increasing much more than it grows with the angle. Notice that we get the largest Rabi splitting for the HPEP propagating perpendicular to the main axis because we set $\sigma_\perp>\sigma_{\parallel}$. If one take $\sigma_\perp<\sigma_{\parallel}$, the opposite behavior will occur. Thus, the most pronounced HPEPs can be excited in the plasmon-exciton hybrid without spacer and along the direction with the highest conductivity. Notice that for the case with $d=0$, in order to homogenize the plasmonic layer separately from the excitonic one, an electrical insulating spacer should be placed between them. However, a few-nm-thick h-BN spacer is enough to insulate layers, thus, in THz range the ratio $d/\lambda$ can be so small that the results obtained from the dispersion relations at $d=0$ will be true. 

\subsection{Plasmonic uniaxial metasurfaces on metal or dielectric films} \label{Subs3b}
\begin{figure*}[htp]
	\centering
	\includegraphics[width=0.94\textwidth]{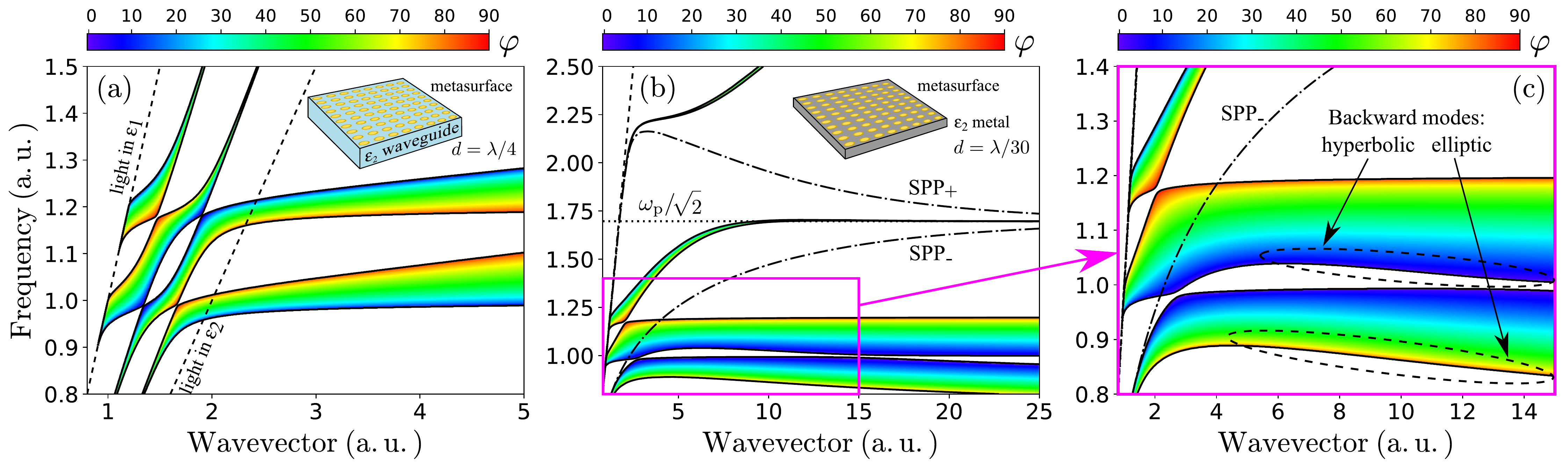} 
	\caption{\label{Fig5} The dispersions $\omega(q)$ of hybrid surface waves at different propagation directions $\varphi$ [see Fig.~\ref{Fig1}(a)] in the plasmonic uniaxial metasurface on the positive-$\varepsilon$ and negative-$\varepsilon$ substrates: the dielectric waveguide with $\varepsilon_2=4,\,\mu_2=1,\,d=\lambda/4$ (a) and thin metal film with $\varepsilon_2=1-\omega_{\rm p}^2/\omega^2,\,\omega_{\rm p}=2\Omega_\perp,\,d=\lambda/30$ (b), respectively, where $\lambda=2\pi c/\omega$. The dashed lines show the dispersion of light in a free space ($\varepsilon_1=1$) and in the waveguide ($\varepsilon_2$). In (b) the dotted line denotes the surface plasmon constant and the dash-dotted lines show the upper and lower SPPs dispersions in the metal substrate. (c) The enlarged fragment of the panel (b), where the dashed circles indicate areas containing backward waves with negative group velocity. All the structures are considered in a free space with $n_1=n_3=1$. The plasmonic metasurface parameters are the same as in Fig.~\ref{Fig2}.}  
\end{figure*}

Here we consider the influence of different nonmagnetic substrates on the surface waves behavior in plasmonic uniaxial metasurfaces. Using Eq.~(\ref{SPPexpl}) with no second 2D layer ($\widehat\sigma_2=0$) and plasmonic layer conductivity $\widehat\sigma_1$ from Fig.~\ref{Fig2}, we calculate at $n_1=n_3=1$ the angle distribution of the surface waves dispersion in the plasmonic uniaxial metasurface on the positive-$\varepsilon$ and negative-$\varepsilon$ substrates. As an example of positive-$\varepsilon$ substrate we considered dielectric WG with $\varepsilon_2=4,\,\mu_2=1,\,d=\lambda/4$ [see Fig.~\ref{Fig5}(a)]. Comparing with Fig.~\ref{Fig3}(e), one can see that the dispersion of surface waves simply shifts to the larger wave vectors $q$, and between the lines of light in WG ($\varepsilon_2$) and out of it ($\varepsilon_1$) a set of hybrid TM-TE WG modes arise with the angle distribution of the dispersion containing similar anticrossings as for the surface waves. The presence of this anticrossings means that there is a mixing of polarization. However, in the case of negative-$\varepsilon$ substrates, the surface waves dispersion can be dramatically changed. As a negative-$\varepsilon$ substrate we took thin metal film with $\varepsilon_2=1-\omega_{\rm p}^2/\omega^2,\,\omega_{\rm p}=2\Omega_\perp,\,d=\lambda/30$. From Fig.~\ref{Fig5}(b) it is seen that the metal substrate gives an additional two sets of hybrid surface waves branches: $\varphi$-dependent, lying below the ordinary branches of a free-standing plasmonic array, and $\varphi$-degenerate, lying above the ordinary branches. At small $q$, the branch $\varphi=0^\circ$ from the lower set coincides with the lower SPP in the considered metal substrate, and the upper $\varphi$-degenerate modes coincide with the upper SPP. Moreover, the presence of the metal substrate also changes the dispersion of the ordinary branches. The upper set of ordinary branches for all angles at high $q$ tends to the surface plasmon constant in the substrate $(\omega_{\rm p}/\sqrt{2})$. The lower set of ordinary branches near $\Omega_\parallel$ gets two anticrossings with the lower set of additional modes: one at small $q\sim2.5$ and another at higher $q\sim15$ [see Fig.~\ref{Fig5}(b)]. Interestingly that both these anticrossings occur at $\varphi\rightarrow0^\circ$, while the ordinary anticrossing, as in a free-standing plasmonic array [Fig.~\ref{Fig3}(e)], corresponds to $\varphi\rightarrow90^\circ$. Moreover, the presence of the second high-$q$ anticrossing results in appearance at $q\sim5-15$ of two sets of backward waves with negative group velocity. The \textit{elliptic backward waves} at $\omega<\Omega_\parallel=1$ exist at high $\varphi$ and the \textit{hyperbolic backward waves} at $\omega>\Omega_\parallel$ can be excited at low $\varphi$ [see Fig.~\ref{Fig5}(c), which is enlarged fragment of Fig.~\ref{Fig5}(b)]. 
\begin{figure*}[htp]
	\centering
	\includegraphics[width=0.94\textwidth]{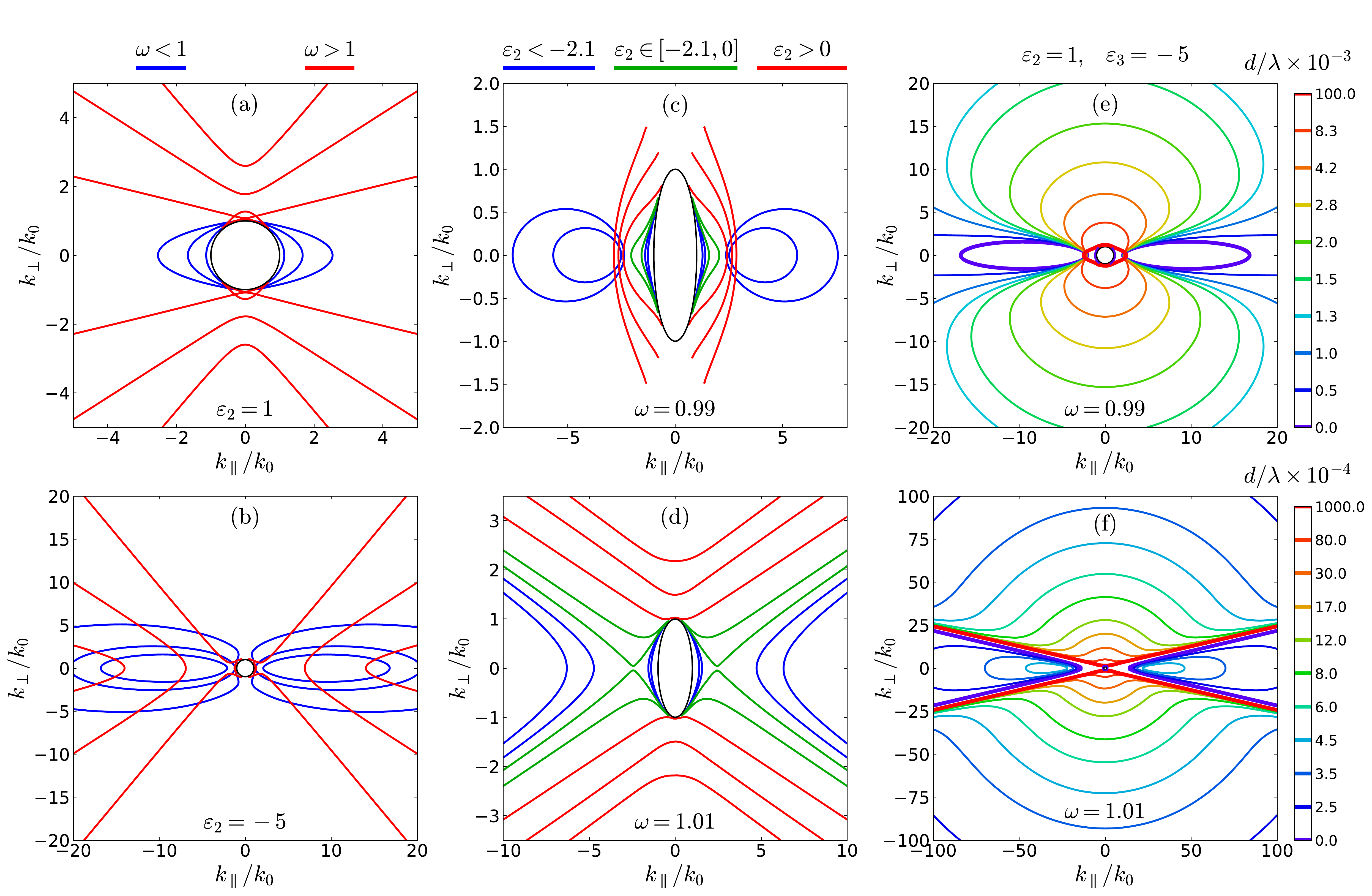} 
	\caption{\label{Fig6} The iso-frequency contours (IFCs) of the hybrid TE-TM surface waves in the plasmonic uniaxial metasurface near the transition frequency of elliptic topology to hyperbolic ($\omega=\Omega_\parallel=1$) in a free space (a) and on the semi-infinite substrate with $\varepsilon_2=-5$ (b). IFCs for different semi-infinite dielectric substrates in the elliptic regime at $\omega=0.99$ (c) and in the hyperbolic one at $\omega=1.01$ (d): additional topological transitions arise in the both regimes when $\varepsilon_2\lesssim-2.1$, in the range $\varepsilon_2\in[-2.1,0]$ the shape of the IFCs changes but the topological transitions do not occur yet. IFCs for the metasurface located at various normalized to $\lambda=2\pi c/\omega$ distances ($d/\lambda$) from the metal semi-infinite substrate with $\varepsilon_3=-5$ in the elliptic regime at $\omega=0.99$ (e) and in the hyperbolic one at $\omega=1.01$ (f). In (e) and (f) the medium between the metasurface and the substrate with $\varepsilon_3$ is a free space with $\varepsilon_2=\varepsilon_1=1$, the bold IFCs correspond to the cases without substrate and on the substrate with zero distance to the metasurface. In all panels the black oval $k_\parallel^2+k_\perp^2=\omega^2/c^2\equiv k_0^2$ denotes the circle of light in a free space and the plasmonic metasurface parameters are the same as in Fig.~\ref{Fig2}.} 
\end{figure*} 

However, in the most curious way, the presence of a negative-$\varepsilon$ substrate is manifested in the topology of IFCs of the hybrid surface waves. The monolayer-like dispersion (\ref{SPP1}) in terms of in-plane wave vectors along the principle axes $k_{\parallel,\perp}$ can be written as:
\begin{widetext}
\vspace{-1em} 
	\begin{align}
		\left[\left(\frac{\kappa_1}{\mu_1}+\frac{\kappa_2}{\mu_2}\right)\left(k_\parallel^2+k_\perp^2\right)-i\left(\sigma_\perp k_\parallel^2 +\sigma_\parallel k_\perp^2\right)\right]\!\!\left[\left(\frac{\varepsilon_1}{\kappa_1}+\frac{\varepsilon_2}{\kappa_2}\right)\left(k_\parallel^2+k_\perp^2\right)+i\left(\sigma_\parallel k_\parallel^2 +\sigma_\perp k_\perp^2\right)\right]=\left(\sigma_\perp-\sigma_\parallel\right)^2k_\parallel^2k_\perp^2,	
	\end{align}
\end{widetext}
where $\kappa_j=\sqrt{(k_\parallel^2+k_\perp^2)\big/k_0^2-\varepsilon_j\mu_j}$, $k_0=\omega/c$, $j=1,2$. Using this expression we numerically found IFCs of the metasurface on different nonmagnetic semi-infinite substrates as an implicit dependency $k_{\perp}(k_\parallel)$ at fixed frequencies. As it was mentioned in Sec.~\ref{Subs3a}, the considered free-standing plasmonic metasurface at $\omega<\Omega_\parallel=1$ possesses the elliptic topology of IFCs and at $\omega>\Omega_\parallel$ the hyperbolic one [see Fig.~\ref{Fig6}(a)]. Note that the arcs arising in the hyperbolic regime near the circle of light in a free space are correspond to the weakly guided quasi-TE modes. Now we obtain that positive-$\varepsilon$ substrates have no fundamental influence on the IFCs, while some negative-$\varepsilon$ substrates result in additional topological transition in the both elliptic and hyperbolic contours [see Fig.~\ref{Fig6}(b)]. The elliptic contours at $\omega=0.99$ and $\varepsilon_2\lesssim-2.1$ are divided into three areas: quasi-TE modes remain arcs around the circle of light and quasi-TM modes become two separate ovals on either side of this circle [see Fig.~\ref{Fig6}(c)]. Here the red arcs are not closed because they are limited by the circles of light corresponding to different $\varepsilon_2>0$. The range $\varepsilon_2\in[-2.1,0]$, when the shape of the contours changes but the topological transition does not occur yet, is caused by the quasi-TE modes contribution in the hybrid surface waves dispersion. The hyperbolic contours at $\omega=1.01$ when $\varepsilon_2$ goes below zero begin to bend in such a way, that at $\varepsilon_2\approx-2.1$, the lower and upper hyperbolas touch each other, and at $\varepsilon_2\lesssim-2.1$ they are split into hyperboles rotated on $90^\circ$ and the elliptic contours near the circle of light [see Fig.~\ref{Fig6}(d)]. For the better description of the considered topological transitions, using Eq.~(\ref{SPPexpl}) we plot in Figs.~\ref{Fig6}(e) and \ref{Fig6}(f) the IFCs of the hybrid surface waves in the metasurface located at various distances ($d$) from the metal semi-infinite substrate. At a very large distance there are ordinary elliptic and hyperbolic contours as in the bare metasurface. When the metal substrate becomes closer to the metasurface, the contours bend in such a way that, finally, at $d\rightarrow0$ they are divided as described above. During this bending the contours pass through the different forms of fourth-order curves \cite{Lawrence}: the elliptic contours are transformed through the forms of a hippopede curve or Cassinian oval, which at $d\rightarrow0$ split into two separate ovals [see Figs.~\ref{Fig6}(e)], and the hyperbolic contours are transformed similar to a so-called Devil's curve [see Figs.~\ref{Fig6}(f)]. Notice that a similar behavior with backward waves and separate ovals in IFCs can be observed in mushroom-type metasurfaces \cite{mushroom}, that also consist of a plasmonic array and a metal ground plane located below it. 

\subsection{Bilayer hyperbolic metasurfaces} \label{Subs3c}
Here we consider bilayer metasurfaces consisting of two 2D layers with effective conductivity tensors $\widehat\sigma_1$ and $\widehat\sigma_2$, which are separated by some interlayer with refractive index $n_2$ and thickness $d$ and surrounded by semi-infinite media with refractive indexes $n_1$ and $n_3$. In the case of the symmetrical bilayer configuration, when $\widehat\sigma_1=\widehat\sigma_2\equiv\widehat\sigma$ and $n_3=n_1$, which gives $P_{23}^{++}=P_{12}^{++}$, $S_{23}^{++}=S_{12}^{++}$, $P_{23}^{--}=-P_{12}^{-+}$, $S_{23}^{--}=-S_{12}^{-+}$, Eq.~(\ref{SPP2}) reduces to the form 
\begin{align}
	\frac{\left(P_{12}^{++}\pm P_{12}^{-+}e^{-\kappa_2d}\right)\left(S_{12}^{++}\pm S_{12}^{-+}e^{-\kappa_2d}\right)}{\left(1\pm e^{-\kappa_2d}\right)^2}=\sigma_{xy}\sigma_{yx}. \label{SPPscreen}
\end{align}
This configuration is equivalent to the monolayer metasurface at the interface between media with $n_1$ and $n_2$ above a perfect mirror on the distance $d/2$, where the perfect mirror condition $n_3\rightarrow\infty$ gives $P_{23}^{--}\big/P_{23}^{++}=S_{23}^{--}\big/S_{23}^{++}=\pm1$, thus, immediately yielding from Eq.~(\ref{SPPexpl}) the same relation as given by Eq.~(\ref{SPPscreen}). This dispersion can be written in a more conventional form
\begin{widetext}
\vspace{-1em} 
	\begin{align}
		\omega_{\rm opt}:\quad\left(\frac{\kappa_1}{\mu_1}+\frac{\kappa_2}{\mu_2}\tanh\left(\kappa_2d/2\right)-i\sigma_{yy}\right)\left(\frac{\varepsilon_1}{\kappa_1}+\frac{\varepsilon_2}{\kappa_2}\tanh\left(\kappa_2d/2\right)+i\sigma_{xx}\right)=\sigma_{xy}\sigma_{yx}, \label{w_o}
		\\[0.5em]
		\omega_{\rm ac}:\quad\left(\frac{\kappa_1}{\mu_1}+\frac{\kappa_2}{\mu_2}\coth\left(\kappa_2d/2\right)-i\sigma_{yy}\right)\left(\frac{\varepsilon_1}{\kappa_1}+\frac{\varepsilon_2}{\kappa_2}\coth\left(\kappa_2d/2\right)+i\sigma_{xx}\right)=\sigma_{xy}\sigma_{yx}.\label{w_a}
	\end{align} 
\end{widetext}
\begin{figure*}[htp]
	\centering
	\includegraphics[width=1\textwidth]{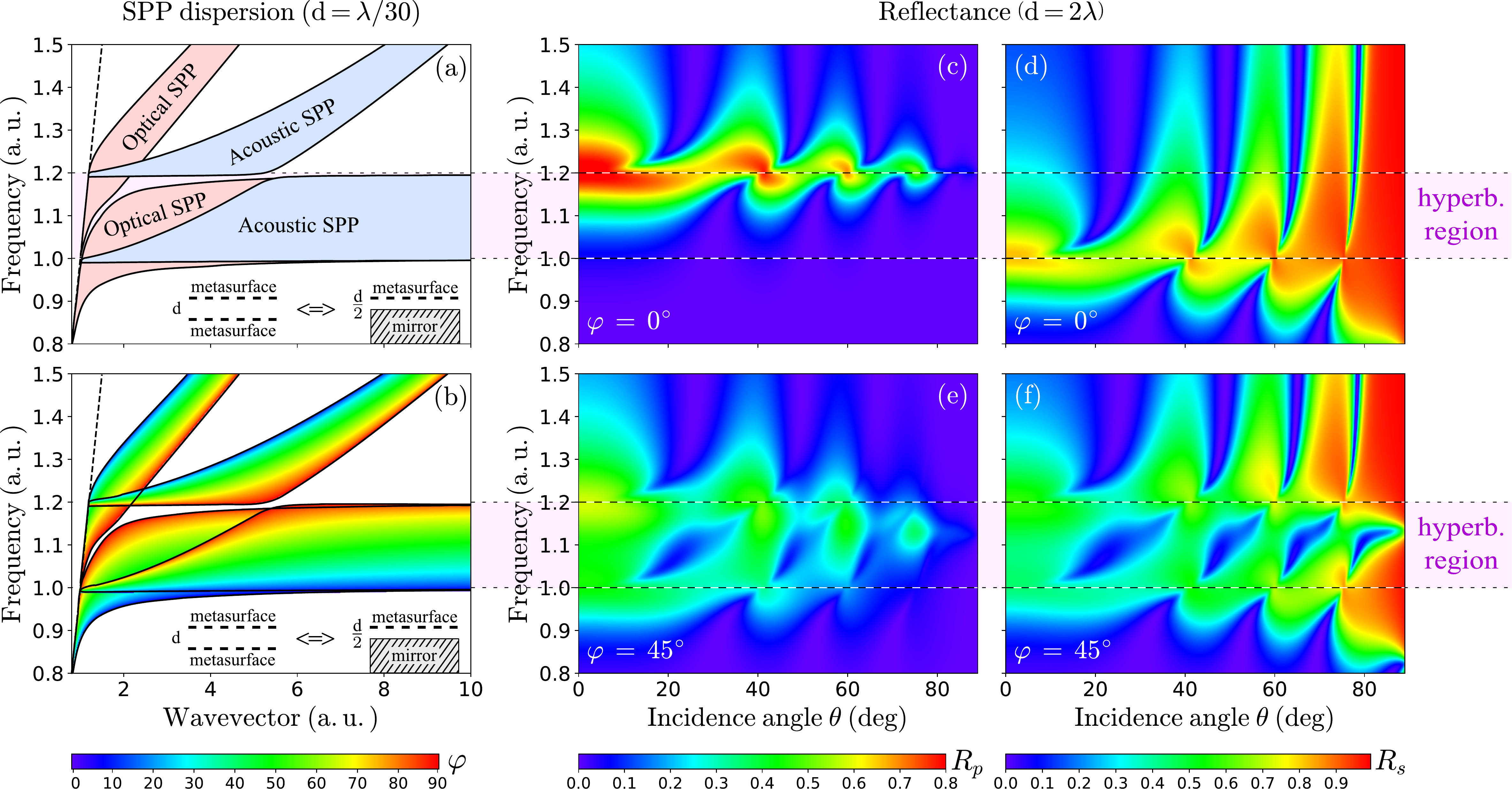} 
	\caption{\label{Fig7} The shaded regions (a) and the propagation angles distribution colormap (b)  of the hybrid surface waves dispersions $\omega(q)$ in the free-standing bilayer metasurface consisting of two identical plasmonic layers with effective conductivities $\widehat\sigma_1=\widehat\sigma_2$ and free-space interlayer ($n_1=n_2=n_3=1,\,d=\lambda/30,\, \lambda=2\pi c/\omega$), or consisting of the free-standing single plasmonic layer above a perfect mirror on the distance $d/2$. (c)-(f) The reflectance $R_{p,s}(\omega,\theta)$, depending on the frequency and angle of incidence $\theta$ [see Fig.~\ref{Fig1}(b)], of the same bilayers but with large free-space interlayer ($n_1=n_2=n_3=1,\, d=2\lambda$) supporting Fabry-Perot resonances. The panels (c)-(d) show $R_{p,s}$ at the in-plane tilted angle $\varphi=0^\circ$, and (e)-(f) at $\varphi=45^\circ$ [see Fig.~\ref{Fig1}(a)]. The reflectance $R_{p}$ for the p-waves shown in (c), (e) and $R_{s}$ for the s-waves in (d), (f). The hyperbolic region lying at $\omega\in[\Omega_{\parallel},\Omega_{\perp}]$ is marked as a unified shaded bar. The plasmonic layers parameters are the same as in Fig.~\ref{Fig2}.}
\end{figure*}
As in any bilayer, the inter-layer EM interaction splits the spectrum into two branches: the solution lying above the dispersion curve of SPPs in a single 2D layer, corresponding to the \textquotedblleft{optical}\textquotedblright\, mode with symmetrical field profile, and a lower frequency solution, corresponding to the \textquotedblleft{acoustic}\textquotedblright\, mode with antisymmetric field profile \cite{GrPl_book}. Here, in Eqs.~(\ref{w_o}) and (\ref{w_a}) for the optical and acoustic branches of the hybrid waves, respectively, we obtained similar bilayer dispersions but with mixed TE-TM terms. In Figs.~\ref{Fig7}(a), (b) we plot the dispersions given by Eqs.~(\ref{w_o}) and (\ref{w_a}) for the free-standing bilayer plasmonic metasurface with a free-space interlayer ($n_1=n_2=n_3=1,\, d=\lambda/30$) at different angles $\varphi$. Comparing with the monolayer dispersion from Fig.~\ref{Fig3}(e), one can see that the optical branches are shifted to the lower $q$ and the acoustic ones to the higher $q$. Thus, at fixed frequency the acoustic modes possess higher $q$, i.e., stronger filed confinement, and at fixed wave vector the optical modes have higher $\omega$ allowing to transfer higher energies. Notice that in contrast to the bilayer with isotropic 2D layers, here at all angles $\varphi\neq0^\circ,\,90^\circ$ we see the intersections of the optical and acoustic branches without any hybridization. These non-hybridized solutions of Eq.~(\ref{SPPscreen}) have a twofold degeneracy, caused by the symmetry of the considered bilayer configuration, and can be excited independently, as they possess different symmetries of filed profiles. Of course, for an asymmetric two-layer configuration, when $\widehat\sigma_1\neq\widehat\sigma_2$ or $n_3\neq n_1$, this degeneracy is removed, and hybridization of the optical and acoustic branches occurs. We underline that the hyperbolic regime in each 2D layer of the considered bilayer metasurface leads to the existence of \textit{hyperbolic acoustic hybrid waves}, which are strongly confined near the bilayer metasurface (out-of-plane confinement) and at the same time possess perfect canalization along some specific directions in the plane of 2D layers  (in-plane confinement). Using Eqs.~(\ref{rrtt}) and (\ref{RRTT}) for the bilayer T-matrix (\ref{T_bi}), in Figs.~\ref{Fig7}(c)-(f) we plot the reflectance $R_{p,s}(\omega,\theta)$, depending on the frequency and angle of incidence [see Fig.~\ref{Fig1}(b)], of the free-standing bilayer plasmonic metasurface with large free-space interlayer ($n_1=n_2=n_3=1,\, d=2\lambda$) supporting Fabry-Perot (FB) resonances. When the plane of incidence of light is at an angle $\varphi=0^\circ$ to the main axis of uniaxial plasmonic layers [see Fig.~\ref{Fig1}(a)], we obtain  FB resonances oscillating with the angle of incidence $\theta$ near the resonant frequencies of plasmonic layers: near $\Omega_{\parallel}=1$ for the s-polarized [Fig.~\ref{Fig7}(c)] and near $\Omega_{\perp}=1.2$ for the p-polarized [Fig.~\ref{Fig7}(d)] incident light. At $\varphi=45^\circ$ an equal mixing of resonant frequencies occurs, and for both polarizations FB resonances completely occupy the region between $\Omega_{\parallel}$ and $\Omega_{\perp}$, so they mostly belong to the hyperbolic regime [see Figs.~\ref{Fig7}(e),(f)]. Notice that all the obtained results for bilayer metasurfaces with the distance $d$ will be the same for a monolayer metasurface above a perfect mirror on the distance $d/2$.  

\subsection{Twisted bilayer hyperbolic metasurfaces} \label{Subs3d}
\begin{figure*}[htp]
	\centering
	\includegraphics[width=1\textwidth]{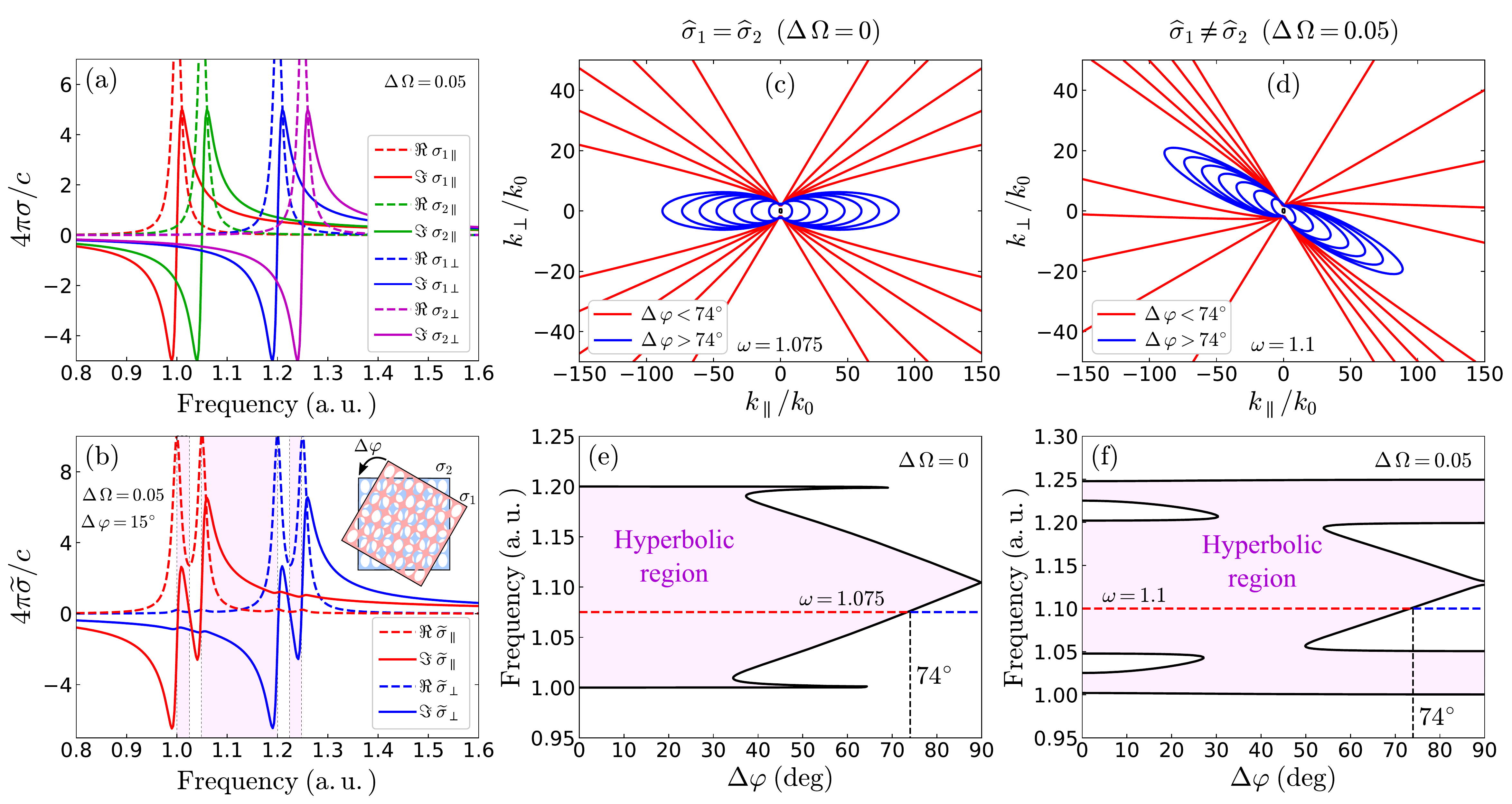} 
	\caption{\label{Fig8} The effective conductivities, iso-frequency contours (IFCs), and  topological transitions diagrams for the thin twisted bilayer metasurface consisting of two 2D uniaxial plasmonic arrays, which are stacked together without spacer ($d=0$) but with a relative in-plane twist angle $\Delta\varphi$. (a) The effective conductivities of the top layer  $\sigma_{1\parallel,\perp}$ with parameters as in Fig.~\ref{Fig2}, and similar bottom layer conductivities $\sigma_{2\parallel,\perp}$ with the detuning $\Delta\Omega=0.05$ of the both resonance frequencies. (b) The effective twisted bilayer conductivities $\widetilde{\sigma}_{\parallel,\perp}$ given by Eq.~(\ref{sigM_add}). (c) and (d) The topological transition of the IFCs to the elliptic behavior at $\Delta\varphi=74^\circ$ on the hyperbolic regime frequencies.  (e) and (f) The diagrams of the IFCs topological transitions versus frequency and twist angle. (c), (e) correspond to zero detuning and (d), (f) to the detuning $\Delta\Omega=0.05$. All the structures are considered in a free space with $\varepsilon=\mu=1$. The shaded regions denote the hyperbolic regime.
	} 
\end{figure*}

Finally, let us consider thin twisted bilayer metasurfaces consisting of two 2D plasmonic layers with a spacer between them and a relative in-plane rotation (set by the twist angle $\Delta\varphi$) [see the inset of Fig.~\ref{Fig8}(b)]. The surface waves dispersions and iso-frequency contours (IFCs) for such a bilayer are given by the general relation (\ref{SPP2}), while being transcendental for the wave vectors, is very complicated for the IFCs topology analyzes. For simplicity, we will consider the bilayer with a spacer, which is thick enough to provide the electrical insulation of the layers but sufficiently thin to neglect the EM filed resonances between them. The electrical insulation permits to homogenize each layer separately, and the thin spacer condition ($d\ll\lambda_{\rm SPP}$) allows the bilayer dispersion (\ref{SPP2}) to be approximated in the zeroth order in $\kappa_2d$ by the monolayer-like relation (\ref{SPPd0}) with the total effective conductivity tensor. Such a reducing of the bilayer problem to a monolayer one significantly simplifies the analysis, although it does not address the case of moir\'{e} metasurfaces \cite{Moire_Rev}, because it does not account for the inter-layer meta-atoms interactions and, therefore, the moir\'{e} superperiod dependence. So, for a thin bilayer described by the monolayer-like Eq.~(\ref{SPPd0}), the effective conductivity tensor can be written as a sum of the respectively rotated conductivity tensors of the top ($\widehat\sigma_1$) and bottom ($\widehat\sigma_2$) layers: 
\vspace{-1em} 
\begin{widetext}
	\vspace{-1em} 
\begin{align}
	&\widehat\sigma_\Sigma=(\widehat\sigma_1)_{\varphi+\Delta\varphi/2}+(\widehat\sigma_2)_{\varphi-\Delta\varphi/2}  \nonumber 
	\\[0.5em]
	&=\begin{pmatrix}
		\sigma_{1\parallel}\cos^2\left(\varphi+\Delta\varphi/2\right) +\sigma_{1\perp}\sin^2\left(\varphi+\Delta\varphi/2\right) &  \left(\sigma_{1\perp} -\sigma_{1\parallel}\right)\sin\left(2\varphi+\Delta\varphi\right)/2 \\[0.5em] \left(\sigma_{1\perp} -\sigma_{1\parallel}\right)\sin\left(2\varphi+\Delta\varphi\right)/2 &  \sigma_{1\parallel}\sin^2\left(\varphi+\Delta\varphi/2\right) +\sigma_{1\perp}\cos^2\left(\varphi+\Delta\varphi/2\right)
	\end{pmatrix} 
	\\[0.5em]
	&+\begin{pmatrix}
		 \sigma_{2\parallel}\cos^2\left(\varphi-\Delta\varphi/2\right) +\sigma_{2\perp}\sin^2\left(\varphi-\Delta\varphi/2\right) &  \left(\sigma_{2\perp} -\sigma_{2\parallel}\right)\sin\left(2\varphi-\Delta\varphi\right)/2 \\[0.5em] \left(\sigma_{2\perp} -\sigma_{2\parallel}\right)\sin\left(2\varphi-\Delta\varphi\right)/2 &  \sigma_{2\parallel}\sin^2\left(\varphi-\Delta\varphi/2\right) +\sigma_{2\perp}\cos^2\left(\varphi-\Delta\varphi/2\right)
	\end{pmatrix}. \nonumber
\end{align}
For convenience we made a symmetrical rotation in the top and bottom layers at the angles $\Delta\varphi/2$ and $-\Delta\varphi/2$, respectively. After some algebra one get 
\begin{align}
	\widehat\sigma_\Sigma=
	\begin{pmatrix}
		\widetilde{\sigma}_{\parallel}\cos^2{\varphi} +\widetilde{\sigma}_{\perp}\sin^2{\varphi}+\delta\sin{\Delta\varphi}\sin{2\varphi} &  \!\!\!\qquad\left(\widetilde{\sigma}_{\perp} -\widetilde{\sigma}_{\parallel}\right)\sin{2\varphi}/2+\delta\sin{\Delta\varphi}\cos{2\varphi}\\[0.5em] \!\!\quad\left(\widetilde{\sigma}_{\perp} -\widetilde{\sigma}_{\parallel}\right)\sin{2\varphi}/2+\delta\sin{\Delta\varphi}\cos{2\varphi} & \!\!\quad\widetilde{\sigma}_{\parallel}\sin^2{\varphi} +\widetilde{\sigma}_{\perp}\cos^2{\varphi}-\delta\sin{\Delta\varphi}\sin{2\varphi}
	\end{pmatrix}, \label{sigM}
\end{align}
where
\begin{align}
	&\widetilde{\sigma}_{\parallel}=\mu_{\parallel}\cos^2\left(\Delta\varphi/2\right) +\mu_{\perp}\sin^2\left(\Delta\varphi/2\right),\quad  \nonumber \widetilde{\sigma}_{\perp}=\mu_{\parallel}\sin^2\left(\Delta\varphi/2\right)+\mu_{\perp}\cos^2\left(\Delta\varphi/2\right),\\ 
	&\widetilde{\sigma}_{\perp}-\widetilde{\sigma}_{\parallel}=\left( \mu_{\perp}-\mu_{\parallel}\right)\cos\left(\Delta\varphi\right),\quad \widehat\mu=\widehat\sigma_1+\widehat\sigma_2,\quad \delta=\frac{\sigma_{1\perp}-\sigma_{2\perp}-\left(\sigma_{1\parallel}-\sigma_{2\parallel}\right)}{2}. \label{sigM_add}
\end{align}

\end{widetext}
\vspace{-1em} 
In general, the rotation at an angle $\varphi$ (measured from the x-axis) of any fully-populated conductivity tensor with $\sigma_{xy}=\sigma_{yx}$ is written as \cite{Alu_OME}:
\vspace{-1em} 
\begin{widetext}
\vspace{-1em}	
\begin{align}
	\begin{pmatrix}
		\sigma_{xx} & \sigma_{xy} \\[0.5em]  \sigma_{xy} & \sigma_{yy}
	\end{pmatrix}_{\varphi} = 
	\begin{pmatrix}
		\sigma_{xx}\cos^2{\varphi} +\sigma_{yy}\sin^2{\varphi} +\sigma_{xy}\sin{2\varphi} &  \qquad\left(\sigma_{yy} -\sigma_{xx}\right)\sin{2\varphi}/2 +\sigma_{xy}\cos{2\varphi} \\[0.5em] \quad\left(\sigma_{yy} -\sigma_{xx}\right)\sin{2\varphi}/2 +\sigma_{xy}\cos{2\varphi}& \quad\sigma_{xx}\sin^2{\varphi} +\sigma_{yy}\cos^2{\varphi}-\sigma_{xy}\sin{2\varphi}
	\end{pmatrix}. \label{tensor}
\end{align}
\end{widetext}
\vspace{-1em} 
Notice that the same sign of the nondiagonal conductivities $\sigma_{xy}=\sigma_{yx}$ corresponds to the systems with lack of inversion symmetry (e.g., with \textit{intrinsic} chirality), while the systems where time-reversal symmetry is broken (e.g., by a magnetic filed) must have $\sigma_{xy}=-\sigma_{yx}$ \cite{Vignale_book}. Comparing Eq.~(\ref{sigM}) with Eq.~(\ref{tensor}), one can rewrite the effective conductivity tensor of the considered twisted bilayer metasurface in the following rotated form  
\begin{align}
	\widehat\sigma_\Sigma=
	\begin{pmatrix}
		\widetilde{\sigma}_{\parallel} &  \delta\sin{\Delta\varphi}\\[0.5em]  \delta\sin{\Delta\varphi}&  \widetilde{\sigma}_{\perp}
	\end{pmatrix}_{\varphi}. \label{sig_twist}
\end{align}
As expected, the layers twist results in the effective chirality response $\sigma_{xy}=\delta\sin{\Delta\varphi}$. For any monolayer-like dispersion (\ref{SPP1}) with the conductivity tensor (\ref{sig}), it can be shown that at large $q$ and $n_1=n_2=1$ the asymptotic behavior of the IFCs of the hybrid surface waves is defined by zeros of $\sigma_{xx}$ (for quasi-TM waves) and $\sigma_{yy}$ (for quasi-TE waves), i.e., by second-order curve equations $\sigma_\parallel k_\parallel^2 +\sigma_\perp k_\perp^2=0$ and $\sigma_\perp k_\parallel^2 +\sigma_\parallel k_\perp^2=0$, which gives the topological transition from the elliptic to hyperbolic regime at  $\im\sigma_{\parallel}\im\sigma_{\perp}=0$ \cite{Alu_PRL}. However, in general, for a fully-populated conductivity tensor given by Eq.~(\ref{tensor}), the asymptotic behavior, e.g., for quasi-TM waves is defined by $\sigma_\parallel k_\parallel^2 +\sigma_\perp k_\perp^2+2\sigma_{xy}k_\parallel k_\perp=0$, so the topological transition occurs at $\im\sigma_\parallel\im\sigma_\perp=\im\sigma_{xy}^2$. In the bilayer tensor (\ref{sigM}) the role of $\sigma_{xy}$ plays the term $\delta\sin{\Delta\varphi}$. As the IFCs of the considered twisted bilayer is given by a monolayer-like dispersion (\ref{SPPd0}) with the total fully-populated effective conductivity tensor (\ref{sigM}), the topological transition in such a bilayer occurs at
\begin{align}
	\im\widetilde{\sigma}_{\parallel}\im\widetilde{\sigma}_{\perp}=\left( \im\delta\sin{\Delta\varphi}\right)^2. \label{IFC}
\end{align}	
In the simple case, when the top and bottom layers have equal conductivities  ($\widehat\sigma_1=\widehat\sigma_2=\widehat\sigma$, $\widehat\mu=2\widehat\sigma$), the detuning $\delta$ is zero, which simplifies Eq.~(\ref{sigM}) to the form  
\begin{align}
	\widehat\sigma_\Sigma=
	\begin{pmatrix}
		\widetilde{\sigma}_{\parallel}\cos^2{\varphi} +\widetilde{\sigma}_{\perp}\sin^2{\varphi} &  \!\!\!\qquad\left(\widetilde{\sigma}_{\perp} -\widetilde{\sigma}_{\parallel}\right)\sin{2\varphi}/2\\[0.5em] \!\!\quad\left(\widetilde{\sigma}_{\perp} -\widetilde{\sigma}_{\parallel}\right)\sin{2\varphi}/2 & \!\!\quad\widetilde{\sigma}_{\parallel}\sin^2{\varphi} +\widetilde{\sigma}_{\perp}\cos^2{\varphi}
	\end{pmatrix}. \label{equal}
\end{align}
Moreover, the topological transition condition (\ref{IFC}) becomes typical $\im\widetilde{\sigma}_{\parallel}\im\widetilde{\sigma}_{\perp}=0$, which at equal conductivities from Eq.~(\ref{sigM_add}) gives  	
\begin{align}
	&\left[\im\sigma_{\parallel}\cos^2\left(\Delta\varphi/2\right) +\im\sigma_{\perp}\sin^2\left(\Delta\varphi/2\right)\right]\nonumber
	\\[0.5em]
	&\times\left[\im\sigma_{\parallel}\sin^2\left(\Delta\varphi/2\right) +\im\sigma_{\perp}\cos^2\left(\Delta\varphi/2\right)\right]=0. \label{IFC0}
\end{align}
Thus, we obtain the topological transition depending not only on the frequency but also on the twist angle $\Delta\varphi$. Using the dispersion (\ref{SPPd0}) and tensor (\ref{equal}) with the same conductivities $\sigma_{\parallel,\perp}$ as in Fig.~\ref{Fig2} but at $\sigma_{\parallel}^\infty=\sigma_{\perp}^\infty=0$, we calculate the IFCs of the hybrid surface waves in a twisted bilayer with $d=0$ and $\widehat\sigma_1=\widehat\sigma_2$ in the hyperbolic regime $\omega\in[\Omega_{\parallel},\Omega_{\perp}]$ at different twist angles $\Delta\varphi$. We obtain that even at hyperbolic regime frequency $\omega=1.075$ for $\Delta\varphi>74^\circ$ the IFCs become elliptic [see Fig.~\ref{Fig8}(c)]. Using the condition (\ref{IFC0}), in Fig.~\ref{Fig8}(e) we plot the diagram of the IFCs topological transitions versus frequency and twist angle, which clarifies this behavior. It is seen that the hyperbolic frequency region narrows with increasing twist angle and at $\Delta\varphi=90^\circ$ it converges to a point. In this point the main axes of the top and bottom layers are mutually orthogonal, so such a bilayer metasurface becomes effectively isotropic with no hyperbolic regime. At $\Delta\varphi=90^\circ$ Eq.~(\ref{IFC0}) yields the condition $\im\sigma_{\parallel}+\im\sigma_{\perp}=0$. Thus, for the conductivities given by Eq.~(\ref{sig_pl}), at $\sigma_{\parallel}^\infty=\sigma_{\perp}^\infty$ and $A_{\parallel}=A_{\perp}$, neglecting the damping, we obtain the frequency of this point to be $\omega_0=\sqrt{(\Omega_\parallel^2+\Omega_\perp^2)\big/2}$. Working at this frequency one can switch the considered bilayer metasurface to the hyperbolic behavior by a slight layers twist near the position with $\Delta\varphi=90^\circ$. So, in this instability point the bilayer can be switched directly from the effective isotropic configuration to the hyperbolic regime. 

In a general case, when the detuning $\delta$ of the top and bottom conductivities is nonzero, the effective chirality response, given by the right part of Eq.~(\ref{IFC}), results in the rotation of the IFCs, which depends both on the twist angle and detuning [see Figs.~\ref{Fig8}(d)]. Here we took for the bottom layer the same conductivities as for the top one but with the detuning $\Delta\Omega=0.05$ of the both resonance frequencies [see Fig.~\ref{Fig8}(a)]. Then the effective twisted bilayer conductivities $\widetilde{\sigma}_{\parallel,\perp}$ given by Eq.~(\ref{sigM_add}) possess the hyperbolic region divided (depending on $\Delta\varphi$) into several bands [see Fig.~\ref{Fig8}(b)]. Using the condition (\ref{IFC}), in Fig.~\ref{Fig8}(f) we plot the diagram of the IFCs topological transitions versus frequency and twist angle for the considered detuning case. For the low angles $\Delta\varphi\lesssim30^\circ$ there are three hyperbolic bands, at the angles $30^\circ\lesssim\Delta\varphi\lesssim50^\circ$ there is a wide single hyperbolic band with $\omega\in[\Omega_{\parallel},\Omega_{\perp}+\Delta\Omega]$, and for the high angles $\Delta\varphi\gtrsim50^\circ$ there are again three hyperbolic bands, but the lower and upper ones have a detuning width $\Delta\Omega=0.05$ and are independent of $\Delta\varphi$, while the middle one, as in the case of zero detuning, narrows with increasing twist angle, although  converging not to a single point but to some narrow frequency window. This window can be found from Eq.~(\ref{IFC}), which at $\Delta\varphi=90^\circ$ gives the condition $\left(\im\sigma_{1\parallel}+\im\sigma_{2\perp}\right)\left(\im\sigma_{1\perp}+\im\sigma_{2\parallel}\right)=0$. For the top layer conductivities $\sigma_{1\parallel,\perp}$ given by Eq.~(\ref{sig_pl}) with the resonance frequencies $\Omega_{\parallel,\perp}$ and $\sigma_{\parallel}^\infty=\sigma_{\perp}^\infty$, $A_{\parallel}=A_{\perp}$, and for the similar bottom layer conductivities $\sigma_{2\parallel,\perp}$  with the resonance frequencies $\Omega_{\parallel,\perp}+\Delta\Omega$, neglecting the damping, we obtain that this window is limited by the frequencies  $\omega_1=\sqrt{(\Omega_\perp^2+(\Omega_\parallel+\Delta\Omega)^2)\big/2}$ and $\omega_2=\sqrt{(\Omega_\parallel^2+(\Omega_\perp+\Delta\Omega)^2)\big/2}$. So, at $\Delta\Omega\ll\Omega_{\parallel,\perp}$ the width of this frequency window is defined by $\omega_2-\omega_1\approx(\Omega_\perp-\Omega_\parallel)\Delta\Omega\Big/\!\sqrt{2(\Omega_\parallel^2+\Omega_\perp^2)}$. 

By solving the exact bilayer dispersion relation (\ref{SPP2}) numerically, we  obtained that the above results correspond to the low-wave-vector (low-$k$) \textquotedblleft{optical}\textquotedblright\, modes in bilayers with  $d/\lambda\lesssim10^{-5}$. In so thin bilayers the IFCs of the \textquotedblleft{acoustic}\textquotedblright\, modes belong to the high-$k$ range. In contrast to the optical IFCs, the acoustic ones have no topological transitions caused by the twist, so they remain hyperboles at hyperbolic regime frequency for any $\Delta\varphi$. With the spacer thickness increasing these acoustic IFCs enters low-$k$ range and start to form with the optical ones different cross-like contours \cite{Tw_Rodin,Tw_Renuka,Tw_Ge}, which may not support the $\Delta\varphi$-transitions. Nevertheless, for $d/\lambda\sim10^{-4}$ the topological transitions predicted in Fig.~\ref{Fig8} are still exist but with small shift in the critical $\Delta\varphi$. As a possible implementation of a bilayer with such a thickness, two rotated graphene strips arrays with 3-nm-thick h-BN spacer (enough to insulate layers) at operation frequency 10 THz ($\lambda=30\,$\textmu m) can be taken. For the considered in Fig.~\ref{Fig8} confinement range $k<150k_0$ ($ k_0=\omega/c$) at $d=3\,\rm nm$ and  $\lambda=30\,$\textmu m the surface waves wavelength is $\lambda_{\rm SPP}>\lambda/150=200\,\rm nm$, so the condition of the monolayer-like approximation $d\ll\lambda_{\rm SPP}$ works well. 

In graphene-based moir\'{e} structures, the appearance of superlattice minibands leads to the twist-angle-dependent van-Hove singularities in the density of states and the additional interband electronic transitions \cite{Tw_Moon,Tw_Stauber2013}. This gives the corresponding dip-peak structure of the optical conductivity and Drude weight, which fractured into several branches of plasmon spectrum \cite{Tw_Stauber2016,Tw_Polini}. By analogy, one can expect similar behavior in the hyperbolic moir\'{e} metasurfaces, where TM-TE waves fractured spectrum probably will make the IFCs topological transitions diagram [see Fig.~\ref{Fig8}(e)] very sliced. So, the predicted $\Delta\varphi$-transitions perhaps will occur only at some certain angles.  

\section{CONCLUSIONS} \label{Sec4}
We have presented a comprehensive analysis of hybrid TM-TE polarized EM waves propagating along different types of few-layer anisotropic metasurfaces. For this we have developed a generalized 4$\times$4 T-matrix formalism allowing us to calculate the linear optical response of arbitrary anisotropic 2D layers accounting for the mixing of EM waves polarizations. Particularly, using this formalism we have analytically obtained a general dispersion relation for an arbitrary bilayer metasurface. We have analyzed the dispersions and IFCs topology of the hybrid waves for various realizations of few-layer anisotropic metasurfaces in the most general form, not specifying a design of constituent 2D layers and describing their optical properties within the effective conductivity approach. Such an approach does not require a specific scale, making the obtained results applicable in different frequency ranges. Having considered four examples of hybrid uniaxial metasurfaces, we have found the following phenomena.

First, we have studied the plasmon-exciton hybrid metasurface, consisting of 2D uniaxial plasmonic array, 2D excitonic layer (e.g., 2D semiconductor or dye molecules), and a thin dielectric spacer between them. For such a metasurface we predict the existence of \textit{hyperbolic plasmon-exciton polaritons}, which are a coherent superposition of the hybrid TM-TE plasmons supported by the uniaxial plasmonic array and the excitons from the excitonic layer. Their dispersion strongly depends both on the propagation direction in the plane of the layers and the spacer thickness. The largest Rabi splitting for this polaritons can be achieved in the hybrid metasurfaces without spacer and at the excitation in the direction with the highest effective conductivity. In contrast to ordinary plasmon-exciton polaritons on a uniform metal \cite{Yang1991,Madrigalmelchor1992,Balci2012,Mortensen} or isotropic plasmonic array \cite{Agarwal,Ramezani,Ding}, here we have demonstrated the possibility of hyperbolic behavior of such a waves. 

Second, we have considered the influence of the positive-$\varepsilon$ (dielectric) and negative-$\varepsilon$ (metal) substrates on the surface waves behavior in plasmonic uniaxial metasurfaces. For a sufficiently thick dielectric substrate we predict a set of hybrid TM-TE WG modes to arise with the angle distribution of the dispersion containing similar anticrossings as for the surface waves. In the case of a metal substrate the situation changes drastically. The additional sets of hybrid surface waves branches arise, among which both \textit{elliptic and hyperbolic backward surface waves} with negative group velocity can exist. Moreover, we predict \textit{additional topological transitions} in both elliptic and hyperbolic IFCs of the hybrid surface waves caused by the presence of a negative-$\varepsilon$ semi-infinite substrate. When the substrate permittivity is less than zero, IFCs begin to bend in such a way, that at some negative value ($\varepsilon_s\approx-2.1$) the elliptic contours split into two separate ovals and arcs near the circle of light in a free space, while the hyperbolic contours split into hyperboles rotated on $90^\circ$ and the elliptic contours near the circle of light. Besides, the controlling parameters of these topological transitions are not only frequency and  substrate permittivity but also the distance to the metal substrate. 

Third, for the bilayer metasurfaces consisting of two 2D uniaxial plasmonic arrays, as well as for the monolayer metasurfaces above a perfect mirror, we predict the existence of \textit{hyperbolic acoustic hybrid waves}, which possess strong confinement in both out-of-plane and in-plane directions. In contrast to acoustic plasmons in isotropic systems \cite{Ac_Koppens,Ac_Lee}, these hyperbolic acoustic waves are not only strongly confined near the bilayer but also perfectly channeled along some specific directions in the plane of 2D layers. 

Finally, we have studied thin twisted bilayer metasurfaces consisting of two 2D plasmonic layers with a spacer between them and a relative in-plane rotation. The spacer is thick enough to provide the electrical insulation of the layers but sufficiently thin to neglect the EM filed resonances between them. This condition allowed us to use the effective conductivity approach to each layer separately and to reduce the bilayer problem to a monolayer one with the total effective conductivity tensor written in a compact form. This formalism significantly simplifies the analysis, although it does not account for the case of moir\'{e} metasurfaces. We have shown that the layers twist angle can be the control parameter for the topological transitions of IFCs of the hybrid surface waves in such bilayers. These topological transitions turned out to be very sensitive to the layers twist: one can switch between hyperbolic and elliptic regimes just by a slight layers twist near the critical angle, which depends on the operating frequency. For the bilayer with equal conductivities of the layers, we have found the instability point, when this bilayer at some frequency can be switched directly from the effective isotropic configuration to the hyperbolic regime by slight divinations from the position with the mutually orthogonal main axes of the layers. Such a topological transition from closed ellipses to open hyperbolas indicates switching from omnidirectional to highly directional surface waves. For the case of different conductivities of the layers (at nonzero conductivity detuning, which can be controlled, e.g., by gate voltage), we predict the appearance of the effective chirality response, which results in the rotation of the IFCs depending both on the twist angle and detuning. The numerical solution of the exact bilayer equation with a spacer of arbitrary thickness has shown that the predicted topological transitions can be found only for the low-$k$ \textquotedblleft{optical}\textquotedblright\, modes in very thin bilayers with $d/\lambda\sim10^{-4}$, which is realistic for the bilayers consisting of graphene strips arrays separated by a few-nm-thick h-BN spacer and working in THz range. For larger distances between layers various cross-like IFCs will arise \cite{Tw_Rodin,Tw_Renuka,Tw_Ge}, which may not contain the predicted effects. However, in hyperbolic moir\'{e} metasurfaces one can still expect the IFCs topological transitions controlled by the twist angle, but the transitions diagram [see Fig.~\ref{Fig8}(e)] for them probably will be very sliced, so the transitions perhaps will occur only at some certain angles.  

We believe that the developed 4$\times$4 T-matrix formalism for arbitrary anisotropic 2D layers may become a useful tool in the calculation of multifunctional few-layer metasurfaces or van der Waals heterostructures with in-plane anisotropy, where it is important to take into account the TM-TE polarization mixing. The predicted hyperbolic hybrid waves and topological transitions in different few-layer anisotropic metasurfaces can be important for the efficient manipulation and control over surface EM waves, which is promising for a number of applications in on-chip optical technologies.

\section*{Acknowledgments}

The authors are grateful to \mbox{Alex Krasnok} and \mbox{Sergey Remizov} for useful discussions. The work was supported by the Russian Science Foundation (Grant 17-12-01393). 


%

\end{document}